\documentstyle[12pt,epsf]{article}
\newlength{\pubnumber} 

\catcode`\@=11
\@addtoreset{equation}{section}

\def\section{\@startsection{section}{1}{\z@}{3.5ex plus 1ex minus .2ex}
 {2.3ex plus .2ex}{\large\bf}}
\def\subsection{\@startsection{subsection}{2}{\z@}{2.3ex plus .2ex}
 {2.3ex plus .2ex}{\bf}}

\begin{document}

\begin{titlepage}
\samepage{
\setcounter{page}{1}
\rightline{UFIFT--HEP--97--23}
\rightline{\tt hep-ph/???????}
\rightline{August 1997}
\vfill
\begin{center}
 {\Large \bf  Toward The Classification of \\
		The Realistic Free Fermionic Models\\}
\vfill
\vfill
 {\large Alon E. Faraggi\footnote{
        E-mail address: faraggi@phys.ufl.edu}\\}
\vspace{.12in}
 {\it           Institute for Fundamental Theory, \\
                Department of Physics, \\
                University of Florida, \\
                Gainesville, FL 32611,
        USA\\}
\end{center}
\vfill
\begin{abstract}
  {\rm
The realistic free fermionic models have had remarkable success in
providing plausible explanations for various properties of the Standard
Model which include the natural appearance of three generations, the
explanation of the heavy top quark mass and the qualitative structure
of the fermion mass spectrum in general, the stability of the proton
and more. These intriguing achievements makes evident the need to
understand the general space of these models. While the number of
possibilities is large, general patterns can be extracted. In this paper
I present a detailed discussion on the construction of the realistic
free fermionic models with the aim of providing some insight into the
basic structures and building blocks that enter the construction. The
role of free phases in the determination of the phenomenology of the
models is discussed in detail. I discuss the connection between the
free phases and mirror symmetry in (2,2) models and the corresponding
symmetries in the case of the (2,0) models. The importance of the free
phases in determining the effective low energy phenomenology is
illustrated in several examples. The classification of the models in
terms of boundary condition selection rules, real world--sheet fermion
pairings, exotic matter states and the hidden sector is discussed.
                 }
\end{abstract}
\smallskip}
\end{titlepage}

\setcounter{footnote}{0}

% ========================= DEFINITIONS ===================================
\def\l{\label}
\def\beq{\begin{equation}}
\def\eeq{\end{equation}}
\def\beqn{\begin{eqnarray}}
\def\eeqn{\end{eqnarray}}

\def\ie{{\it i.e.}}
\def\eg{{\it e.g.}}
\def\half{{\textstyle{1\over 2}}}
\def\third{{\textstyle {1\over3}}}
\def\quarter{{\textstyle {1\over4}}}
\def\m{{\tt -}}
\def\p{{\tt +}}

\def\slash#1{#1\hskip-6pt/\hskip6pt}
\def\slk{\slash{k}}
\def\GeV{\,{\rm GeV}}
\def\TeV{\,{\rm TeV}}
\def\y{\,{\rm y}}
\def\SM{Standard-Model }
\def\SUSY{supersymmetry }
\def\SSSM{supersymmetric standard model}
\def\vev#1{\left\langle #1\right\rangle}
\def\l{\langle}
\def\r{\rangle}

\def\Htw{{\tilde H}}
\def\chibar{{\overline{\chi}}}
\def\qbar{{\overline{q}}}
\def\ibar{{\overline{\imath}}}
\def\jbar{{\overline{\jmath}}}
\def\Hbar{{\overline{H}}}
\def\Qbar{{\overline{Q}}}
\def\abar{{\overline{a}}}
\def\alphabar{{\overline{\alpha}}}
\def\betabar{{\overline{\beta}}}
\def\tautwo{{ \tau_2 }}
\def\thetatwo{{ \vartheta_2 }}
\def\thetathree{{ \vartheta_3 }}
\def\thetafour{{ \vartheta_4 }}
\def\ttwo{{\vartheta_2}}
\def\tthree{{\vartheta_3}}
\def\tfour{{\vartheta_4}}
\def\ti{{\vartheta_i}}
\def\tj{{\vartheta_j}}
\def\tk{{\vartheta_k}}
\def\calF{{\cal F}}
\def\smallmatrix#1#2#3#4{{ {{#1}~{#2}\choose{#3}~{#4}} }}
\def\ab{{\alpha\beta}}
\def\Minv{{ (M^{-1}_\ab)_{ij} }}
\def\bone{{\bf 1}}
\def\ii{{(i)}}
\def\V{{\bf V}}
\def\b{{\bf b}}
\def\N{{\bf N}}
\def\t#1#2{{ \Theta\left\lbrack \matrix{ {#1}\cr {#2}\cr }\right\rbrack }}
\def\C#1#2{{ C\left\lbrack \matrix{ {#1}\cr {#2}\cr }\right\rbrack }}
\def\tp#1#2{{ \Theta'\left\lbrack \matrix{ {#1}\cr {#2}\cr }\right\rbrack }}
\def\tpp#1#2{{ \Theta''\left\lbrack \matrix{ {#1}\cr {#2}\cr }\right\rbrack }}
\def\l{\langle}
\def\r{\rangle}

%================== BLACKBOARD BOLD CHARACTERS ==============================

\def\inbar{\,\vrule height1.5ex width.4pt depth0pt}

\def\IC{\relax\hbox{$\inbar\kern-.3em{\rm C}$}}
\def\IQ{\relax\hbox{$\inbar\kern-.3em{\rm Q}$}}
\def\IR{\relax{\rm I\kern-.18em R}}
 \font\cmss=cmss10 \font\cmsss=cmss10 at 7pt
\def\IZ{\relax\ifmmode\mathchoice
 {\hbox{\cmss Z\kern-.4em Z}}{\hbox{\cmss Z\kern-.4em Z}}
 {\lower.9pt\hbox{\cmsss Z\kern-.4em Z}}
 {\lower1.2pt\hbox{\cmsss Z\kern-.4em Z}}\else{\cmss Z\kern-.4em Z}\fi}

%========================================================================
%          MACROS FOR REFERENCES
%========================================================================
\def\AEF{A.E. Faraggi}
\def\NPB#1#2#3{{\it Nucl.\ Phys.}\/ {\bf B#1} (19#2) #3}
\def\PLB#1#2#3{{\it Phys.\ Lett.}\/ {\bf B#1} (19#2) #3}
\def\PRD#1#2#3{{\it Phys.\ Rev.}\/ {\bf D#1} (19#2) #3}
\def\PRL#1#2#3{{\it Phys.\ Rev.\ Lett.}\/ {\bf #1} (19#2) #3}
\def\PRT#1#2#3{{\it Phys.\ Rep.}\/ {\bf#1} (19#2) #3}
\def\MODA#1#2#3{{\it Mod.\ Phys.\ Lett.}\/ {\bf A#1} (19#2) #3}
\def\IJMP#1#2#3{{\it Int.\ J.\ Mod.\ Phys.}\/ {\bf A#1} (19#2) #3}
\def\nuvc#1#2#3{{\it Nuovo Cimento}\/ {\bf #1A} (#2) #3}
\def\etal{{\it et al\/}}

%==============================================================================
\hyphenation{su-per-sym-met-ric non-su-per-sym-met-ric}
\hyphenation{space-time-super-sym-met-ric}
\hyphenation{mod-u-lar mod-u-lar--in-var-i-ant}
%==============================================================================

%============================== SECTION 1 ============================

\setcounter{footnote}{0}
\section{Introduction}

Despite interesting progress over the last few years in understanding
non perturbative aspects of superstring theory, trying to
connect string theory to experimental low energy physics
still relies on the analysis of perturbative
heterotic string \cite{HETE} vacua\footnote{For discussions
of phenomenological aspects related to M and F theory see {\it e.g.}
ref. \cite{witten,mfphenomenology}}.
The study of perturbative string vacua has progressed significantly
over the past decade and we now have several
distinct, but perhaps related, ways to construct
perturbative string models in four dimensions.
Among those are the geometrical formulations
that include the Calabi--Yau compactifications \cite{CY}
and related potential constructions \cite{potcon}, and the
orbifold compactifications \cite{DHVW}. On the other side
we have algebraic constructions that include
for example the free fermionic formulation \cite{fff}
and more general constructions obtained by
tensoring higher level conformal field theories \cite{gepner,twozero}.
Similarly we have broaden our view on the possible
open avenues for trying to connect between
string vacua and low energy physics. Among those are
the string GUT models \cite{stringguts}
and in particular the three generation
string GUT models \cite{kakushadze}, the semi--simple
string GUT models \cite{ssimpleguts,revamp,search,alr} and
the superstring standard--like models \cite{z3slm,eu,slm},
in which the non Abelian content of the Standard Model
gauge group is obtained directly at the string level.

Among the semi--realistic superstring models
constructed to date the models in the free
fermionic formulation have gone the furthest
in trying to recapture the physics of the
Standard Model.
Some general properties of this class of models
suggest that their success is not accidental.
First is the fact that the free fermionic construction
is formulated at a highly symmetric point
in the string moduli space. Second, the realistic free
fermionic models correspond to $Z_2\times Z_2$ orbifold
compactifications at an enhanced symmetry point in
the Narain moduli space \cite{NARAIN}.
The existence of three generations
in the free fermionic models is correlated with the
underlying orbifold structure. This is exhibited
in the fact that the free fermionic models produce a
large class of three generation models with varied
phenomenological properties.
The emergence of three generations in these models
is not ad hoc and motivates the hypothesis
that the true string vacuum is a $Z_2\times Z_2$
orbifold at the vicinity of the free fermionic
point in the string moduli space. Another important
property of the three generation free fermionic models
is the fact that the weak hypercharge has the standard
$SO(10)$ embedding. Thus, although the $SO(10)$ symmetry
is broken at the string level rather than in the effective
low energy field theory, some of the appealing features
of standard $SO(10)$ unification are retained while
some of the generic difficulties associated with
grand unification, like doublet--triplet splitting,
may be resolved.

The realistic superstring models \cite{revamp,fny,alr,eu,top,slm}
in the free fermionic formulation \cite{fff}
have had intriguing success in describing the real world.
Some of the important properties of these models are:
\begin{itemize}
\item
The existence of three
generations arises due to the
underlying $Z_2\times Z_2$ orbifold compactification \cite{ztwo}.
Qualitatively realistic fermion mass spectrum can be
envisioned \cite{ffm,nrt,gmh,ckm}.
The generation mass hierarchy and the general texture of the
fermion mass matrices are also seen to arise from the symmetries
of the underlying $Z_2\times Z_2$ orbifold compactification
\cite{gmh}.
\item
The heavy generation Yukawa couplings
have been calculated explicitly in specific models and found
to be in agreement with the observed masses.
The heaviness of the top quark arises because only the top quark
gets a cubic level mass term while the lighter fermion obtain their
mass terms from nonrenormalizable terms \cite{top}.
\item
Dimension four and five operators that may
mediate rapid proton decay are suppressed
due to stringy symmetries and due to a superstring doublet--triplet
splitting mechanism in which the color triplets are projected out
by the GSO projections \cite{ps}. Thus, the superstring derived standard--like
models may naturally give rise to a stable proton \cite{ps,pati}.
\item
Exotic matter states arise in these models due to the
breaking of the non--Abelian grand unified gauge symmetries
at the string level by Wilson lines.
These exotic matter states may yield new dark matter candidates
and new exotic leptoquark states that in general do not arise
in Grand Unified theories \cite{ccf}.
Such states are referred to generically as Wilsonian matter states.
\item
A see--saw mechanism which suppresses left--handed neutrino masses
was proposed \cite{atr,fhneut}. The superstring see--saw mechanism
makes use of the exotic Wilsonian matter states. The neutrino--Higgsino
mixing problem, which is expected to arise generically in Gravity
Unified Theories can also be resolved due to the ``non--standard''
charges of the exotic Wilsonian matter states \cite{fpnhmix}.
\item
An important property of the realistic free fermionic models
is the standard $SO(10)$ embedding of the weak hypercharge.
Consequently, it was suggested that in these models
string gauge coupling unification may be in agreement with
$\sin^2\theta_W(M_Z)$ and $\alpha_s(M_Z)$, provided that there exist
additional matter beyond the Minimal Supersymmetric Standard Model
spectrum at intermediate energy thresholds \cite{price,gcu,dienesfaraggi}.
\end{itemize}
\bigskip
The list of remarkable accomplishments of the realistic free fermionic
models makes evident the need to better understand their general
structure. To date the study of these models has mainly focused on several
isolated examples and a systematic classification
is still lacking. The large space of three generation models
offers on the one hand the intriguing possibility
that there exist a model in this space which satisfies
all of the experimental constraints. On the other hand
this richness makes a systematic classification a seemingly
impossible challenge. The aim of this paper is to provide
some of the insight that has been developed into the
basic structures and building blocks that enter the
construction, thus taking the first steps toward a
classification of the realistic free fermionic models. From
a different view such a classification is required
if we are to uncover whether a fully realistic model
can, or cannot, be obtained in this class of models.

Models in the free fermionic formulation are constructed by specifying
a set of boundary condition basis vectors and a choice of GSO
projection coefficients, consistent with the string consistency constraint.
The low energy spectrum and allowed interactions are then determined.
The classification of the models can then proceed by developing
general rules of how the low energy properties of the models
are determined by the fermionic boundary conditions and
the GSO phases.
To date, some rules based on the world--sheet fermion boundary conditions
have been obtained. However, a general discussion of the basic
building blocks and
of the role of the GSO phases is still lacking.
In this paper I make a small step in trying to fill this gap.
Due to the large number of possibilities the discussion is necessarily
incomplete.
Several general properties, such as mirror
symmetry for (2,2) as well as (2,0) are explained in terms of
free phases.
Furthermore, the notion of duality symmetries
due to the discrete choices of free phases can be
extended to the set of basis vectors and GSO phases
which span the realistic models, thus providing one
mean of classifying the models.
The choices of GSO phases also play an important role
in the determination of the low energy spectrum and the
effective low energy phenomenology.
For example, as will be shown, discrete choices of phases
affect the massless matter states, in a way which affects
the string gauge coupling unification problem.
The classification of the models in terms of the real world--sheet
fermion pairings, the exotic matter states and the hidden sector
is discussed. In several examples it will be shown
how the different choices of real fermion pairing relate
to the phenomenological properties of the models. The classification
of exotics is done by classifying all the possible types
of exotic states that may appear in the models as well
as by the type of exotic states that actually appear in
specific models.
Some phenomenological implications of the
classification by the exotic states and hidden sector are
examined.

The paper is organized as follows. Section 2 contains a
review of the realistic free fermionic construction.
Section 3 contains a discussion of the classification of
the models by the boundary condition rules.
In section 4 the role of the free phases is examined.
The relation between the GSO phases and mirror symmetry
is shown for (2,2) models. It is then shown
that mirror symmetry, as a result of the free phases
also exists in the (2,0) models, interchanging $16$
of $SO(10)$ with the ${\bar{16}}$. The connection
between the free phases and other properties of
the models, like space--time supersymmetry and the
presence of exotics is further discussed.
In section 6, 7 and 8 the classification of the models
in terms of,  the real world--sheet fermion pairings;
the types of exotics; and the hidden sector; is discussed.
Finally, section 9 contains the conclusions.

\setcounter{footnote}{0}
\section{Realistic free fermionic models}

In the free fermionic formulation of the heterotic string
in four dimensions all the world--sheet
degrees of freedom  required to cancel
the conformal anomaly are represented in terms of free fermions
propagating on the string world--sheet.
In the light--cone gauge the world--sheet field content consists
of two transverse left-- and right--moving space--time coordinate bosons,
$X_{1,2}^\mu$ and ${\bar X}_{1,2}^\mu$,
and their left--moving fermionic superpartners $\psi^\mu_{1,2}$,
and additional 62 purely internal
Majorana--Weyl fermions, of which 18 are left--moving,
$\chi^{I}$, and 44 are right--moving, $\phi^a$.
In the supersymmetric sector the world--sheet supersymmetry is realized
non--linearly and the world--sheet supercurrent is given by
\begin{equation}
T_F=\psi^\mu\partial X_\mu+f_{IJK}\chi^I\chi^J\chi^K,
\label{supercurrent}
\end{equation}
where $f_{IJK}$ are the structure constants of a semi--simple
Lie group of dimension 18. The $\chi^{I}~(I=1,\cdots,18)$
world--sheet fermions transform in the adjoint representation of
the Lie group. In the realistic free fermionic models the Lie group is
$SU(2)^6$. The $\chi^I~{I=1,\cdots,18}$ transform in the adjoint
representation of $SU(2)^6$, and are denoted by
$\chi^I,~y^I,~\omega^I~(I=1,\cdots,6)$.
Under parallel transport around a noncontractible loop on the toroidal
world--sheet the fermionic fields pick up a phase
\begin{equation}
f~\rightarrow~-{\rm e}^{i\pi\alpha(f)}f~.
\label{fermionphase}
\end{equation}
The minus sign is conventional and $\alpha(f)\in(-1,+1]$.
Each set of specified
phases for all world--sheet fermions, around all the non--contractible
loops is called the spin structure of the model. Such spin structures
are usually given is the form of 64 dimensional boundary condition vectors,
with each element of the vector specifying the phase of the corresponding
world--sheet fermion. The partition function is a sum over all
such spin structures:
\begin{equation}
    Z(\tau) ~=~  \sum_{\alpha}
      (-1)^F\, {\rm Tr}\,({\alpha})~.
\label{genform}
\end{equation}
where the sum is over all the sectors (spin structures) $\alpha$ of the
theory, $(-1)^F$ is a space--time fermion number operator, and
${\rm Tr}\,({\alpha})$ indicate a trace over the Fock space of mode
excitations of the world--sheet fields. In string models this trace
is generically realized as GSO projections between subsetcors of
the theory,
\begin{equation}
        {\rm Tr}\,(\alpha)~=~ {1\over g}\,\sum_{\beta}
      \,c{\alpha\choose\beta}\, {\rm Tr}\,{{\alpha}\choose{\beta}}~.
\label{GSO}
\end{equation}
Here the $\beta$-sum
implements the GSO projection,
$c{\alpha\choose\beta}$ are the chosen GSO phases,
$g$ is a normalization factor,
and ${\rm Tr}{\alpha\choose\beta}$ indicates a restricted trace
over the appropriate $\alpha\choose\beta$ subsector.

The partition function is required to be invariant under modular
transformation. Modular transformation, in general, mix between
the different spin--structures. Therefore, requiring invariance under
modular transformation results in a set of rules which
constrains the allowed spin--structures and their amplitudes
$c{\alpha\choose\beta}$. It turns out that a string model
can be specified in terms of boundary condition basis vectors
and the GSO projection coefficients for these basis vectors.
The modular invariance constraints are in turn translated to
a set of rules which constrain the allowed boundary condition
basis vectors and their GSO projection coefficients. These rules
are given in ref. \cite{fff}.
In general the partition function is summed over
all the genus--$g$ string world--sheet. Due to factorization theorem
the genus--$g$ amplitude factorizes into a product of genus one amplitudes.
One--loop modular invariance is therefore sufficient
to constrain the consistent spin structures. Thus,
a model in this construction
is defined by a set of boundary conditions basis vectors
and by a choice of generalized GSO projection coefficients, which
satisfy the one--loop modular invariance constraints. The boundary
conditions basis vectors ${\bf b}_k$ span a finite additive group
\begin{equation}
\Xi={\sum_k}n_i {b}_i
\label{additivegroup}
\end{equation}
where $n_i=0,\cdots,{{N_{z_i}}-1}$.
The physical massless states in the Hilbert space of a given sector
$\alpha\in{\Xi}$ are then obtained by acting on the vacuum state of
that sector with the world-sheet bosonic and fermionic mode operators,
with frequencies $\nu_f$, $\nu_{f^*}$ and
by subsequently applying the generalized GSO projections,
\begin{equation}
\left\{e^{i\pi({b_i}F_\alpha)}-
{\delta_\alpha}c^*\left(\matrix{\alpha\cr
                 b_i\cr}\right)\right\}\vert{s}\rangle=0
\label{gsoprojections}
\end{equation}
with
\begin{equation}
(b_i{F_\alpha})\equiv\{\sum_{real+complex\atop{left}}-
\sum_{real+complex\atop{right}}\}(b_i(f)F_\alpha(f)),
\label{lorentzproduct}
\end{equation}
where $F_\alpha(f)$ is a fermion number operator counting each mode of
$f$ once (and if $f$ is complex, $f^*$ minus once). For periodic
complex fermions [{\it i.e.} for $\alpha(f)=1)$]
the vacuum is a spinor in order to represent
the Clifford algebra of the corresponding zero modes.
For each periodic complex fermion $f$,
there are two degenerate vacua $\vert{+}\rangle$, $\vert{-}\rangle$,
annihilated by the zero modes $f_0$ and $f^*_0$ and with fermion
number $F(f)=0,-1$ respectively. In Eq. (\ref{gsoprojections}),
$\delta_\alpha=-1$ if $\psi^\mu$ is periodic in the sector $\alpha$,
and $\delta_\alpha=+1$ if $\psi^\mu$ is antiperiodic in the sector $\alpha$.
The states satisfy the Virasoro condition:
\beq
M_L^2=-{1\over 2}+{{{\alpha_L}\cdot{\alpha_L}}\over 8}+N_L=-1+
{{{\alpha_R}\cdot{\alpha_R}}\over 8}+N_R=M_R^2
\label{virasorocond}
\eeq
where
$\alpha=(\alpha_L;\alpha_R)\in\Xi$ is a sector in the additive group, and
\beq
N_L=\sum_f ({\nu_L}) ;\hskip 3cm N_R=\sum_f ({\nu_R})
\label{nlnr}
\eeq
\beq
\nu_f={{1+\alpha(f)}\over 2} ;\hskip 3cm
{\nu_{f^*}}={{1-\alpha(f)}\over 2}.
\label{nulnur}
\eeq
The $U(1)$ charges with respect to the unbroken Cartan generators of the
four dimensional gauge group are in one to one correspondence with the
$U(1)$ $ff^*$ currents. For each complex fermion $f$:
\begin{equation}
Q(f)={1\over2}\alpha(f)+F(f).
\label{u1numbers}
\end{equation}
The representation (\ref{u1numbers})
shows that $Q(f)$ is identical with the world--sheet
fermion numbers $F(f)$ for states in the Neveu--Schwarz sector
$(\alpha(f)=0)$
and is $(F(f)+{1\over2})$ for states in the Ramond sector
$(\alpha(f)=1)$.
The charges for the $\vert\pm\rangle$ spinor vacua are $\pm{1\over2}$.

\subsection{The NAHE set}
The boundary condition basis vectors which generate the realistic
free fermionic models are, in general, divided into two major subsets.
The first set consist of the NAHE set \cite{nahe,slm}, which is a set
of five boundary condition basis vectors denoted $\{{\bf 1},S,b_1,b_2,b_3\}$.
With `0' indicating Neveu-Schwarz boundary conditions
and `1' indicating Ramond boundary conditions, these vectors are as follows:
\beqn
 &&\begin{tabular}{c|c|ccc|c|ccc|c}
 ~ & $\psi^\mu$ & $\chi^{12}$ & $\chi^{34}$ & $\chi^{56}$ &
        $\bar{\psi}^{1,...,5} $ &
        $\bar{\eta}^1 $&
        $\bar{\eta}^2 $&
        $\bar{\eta}^3 $&
        $\bar{\phi}^{1,...,8} $ \\
\hline
\hline
      {\bf 1} &  1 & 1&1&1 & 1,...,1 & 1 & 1 & 1 & 1,...,1 \\
          $S$ &  1 & 1&1&1 & 0,...,0 & 0 & 0 & 0 & 0,...,0 \\
\hline
  ${b}_1$ &  1 & 1&0&0 & 1,...,1 & 1 & 0 & 0 & 0,...,0 \\
  ${b}_2$ &  1 & 0&1&0 & 1,...,1 & 0 & 1 & 0 & 0,...,0 \\
  ${b}_3$ &  1 & 0&0&1 & 1,...,1 & 0 & 0 & 1 & 0,...,0 \\
\end{tabular}
   \nonumber\\
   ~  &&  ~ \nonumber\\
   ~  &&  ~ \nonumber\\
     &&\begin{tabular}{c|cc|cc|cc}
 ~&      $y^{3,...,6}$  &
        $\bar{y}^{3,...,6}$  &
        $y^{1,2},\omega^{5,6}$  &
        $\bar{y}^{1,2},\bar{\omega}^{5,6}$  &
        $\omega^{1,...,4}$  &
        $\bar{\omega}^{1,...,4}$   \\
\hline
\hline
    {\bf 1} & 1,...,1 & 1,...,1 & 1,...,1 & 1,...,1 & 1,...,1 & 1,...,1 \\
    $S$     & 0,...,0 & 0,...,0 & 0,...,0 & 0,...,0 & 0,...,0 & 0,...,0 \\
\hline
${b}_1$ & 1,...,1 & 1,...,1 & 0,...,0 & 0,...,0 & 0,...,0 & 0,...,0 \\
${b}_2$ & 0,...,0 & 0,...,0 & 1,...,1 & 1,...,1 & 0,...,0 & 0,...,0 \\
${b}_3$ & 0,...,0 & 0,...,0 & 0,...,0 & 0,...,0 & 1,...,1 & 1,...,1 \\
\end{tabular}
\label{nahe}
\eeqn
with the following
choice of phases which define how the generalized GSO projections are to
be performed in each sector of the theory:
\beq
      C\left( \matrix{b_i\cr b_j\cr}\right)~=~
      C\left( \matrix{b_i\cr S\cr}\right) ~=~
      C\left( \matrix{\bone \cr \bone \cr}\right) ~= ~ -1~.
\label{nahephases}
\eeq
The remaining projection phases can be determined from those above through
the self-consistency constraints.
The precise rules governing the choices of such vectors and phases, as well
as the procedures for generating the corresponding space-time particle
spectrum, are given in Refs.~\cite{fff}.

The basis vector $S$ generated the space--time supersymmetry.
The set of basis vectors $\{{\bf 1},S\}$ generates a model
with $N=4$ space--time supersymmetry and $SO(44)$ gauge group
in the right--moving sector. Imposing the GSO projections
of the basis vectors $b_1$, $b_2$ and $b_3$ reduces the $N=4$
supersymmetry to $N=1$ and brakes the gauge group to
$SO(10)\times SO(6)^3\times E_8$.
The space-time vector bosons that generate the gauge group
arise from the Neveu-Schwarz sector and from the sector
$I \equiv {\bf 1}+b_1+b_2+b_3$.
The Neveu-Schwarz sector produces the generators of
$SO(10)\times SO(6)^3\times SO(16)$. The sector ${\bf 1}+b_1+b_2+b_3$
produces the spinorial {\bf 128} of $SO(16)$ and completes the hidden
gauge group to $E_8$.
The three basis vectors $b_1$, $b_2$ and $b_3$ correspond to the three
twisted sectors of the $Z_2\times Z_2$ orbifold. Each one of these sectors
produces 16 multiplets in the ${\bf 16}$ representation of $SO(10)$, .

The correspondence of the NAHE set with $Z_2\times Z_2$ orbifold
is illustrated by adding to the NAHE the boundary condition basis
vector $X$ with periodic boundary conditions for the world--sheet
fermions $\{{\bar\psi}^{1,\cdots,5},{\bar\eta}^{1,2,3}\}$ and all
the other world--sheet fermions have antiperiodic boundary conditions,
\beq
X=(0,\cdots,0\vert{\underbrace{1,\cdots,1}_{{\psi^{1,\cdots,5}},
{\eta^{1,2,3}}}},0,\cdots,0)~.
\label{vectorx}
\eeq
The choice of generalized GSO projection coefficients is
\begin{equation}
      C\left( \matrix{X\cr \b_j\cr}\right)~=~
      -C\left( \matrix{X\cr S\cr}\right) ~=~
      C\left( \matrix{X\cr \bone \cr}\right) ~= ~ +1~.
\label{Xphases}
\end{equation}
With the set $\{{\bf 1},S,b_1,b_2,b_3,X\}$ the gauge group
is $E_6\times SO(4)^3\times E_8$ with 24 generations in the
27 representation of $E_6$. The same model is constructed in the
orbifold formulation by first constructing the Narain lattice
with $SO(12)\times E_8\times E_8$ gauge group $N=4$ supersymmetry.
The gauge group is broken to $E_6\times SO(4)^3\times E_8$
after applying the $Z_2\times Z_2$ twisting on the $SO(12)$ lattice.
The three twisted sectors produce 48 fixed points, which correspond
to the 24 generations in the fermionic construction.

The focus in this paper is on models that contain the full NAHE
set of boundary condition basis vectors. Three generation free
fermionic models that do not use the full NAHE set can also be
constructed \cite{KS,lykken}. Such models correspond to
$Z_2\times Z_2$ orbifold compactifications without the
standard embedding of the gauge connection. However, the NAHE
set facilitates the illustration of the basic ingredients
that enter the construction of the realistic free fermionic
models. These basic ingredients can then also be extended
to models that do not use the full NAHE set.

Note that the NAHE set has an $S_3$ permutation symmetry
obtained by permuting the three basis vectors $b_j$ $(j=1,2,3)$.
This permutation symmetry if again a reflection of the
underlying $Z_2\times Z_2$ orbifold compactification.
The basis vectors which extend the NAHE set will typically
break the permutation symmetry.

\subsection{Beyond the NAHE set}

At the level of the NAHE set the observable gauge group is
$SO(10)\times SO(6)^3$ and the number of generations is 48, sixteen
from each sector $b_1$, $b_2$ and $b_3$. The $SO(6)^3$ symmetries
are horizontal flavor dependent symmetries. To break the gauge
group and to reduce the number of generations, we must add additional
boundary condition basis vectors to the NAHE set. These additional
vectors break the $SO(10)$ and the flavor $SO(6)$ gauge symmetries and
in turn reduce the number of generations to three generations.
The $SO(10)$ gauge group is broken to one of its subgroups
$SU(5)\times U(1)$, $SO(6)\times SO(4)$ or
$SU(3)\times SU(2)\times U(1)^2$ by the assignment of
boundary conditions to the set ${\bar\psi}^{1\cdots5}_{1\over2}$:

1. $b\{{{\bar\psi}_{1\over2}^{1\cdots5}}\}=
\{{1\over2}{1\over2}{1\over2}{1\over2}
{1\over2}\}\Rightarrow SU(5)\times U(1)$,

2. $b\{{{\bar\psi}_{1\over2}^{1\cdots5}}\}=\{1 1 1 0 0\}
  \Rightarrow SO(6)\times SO(4)$.

To break the $SO(10)$ symmetry to
$SU(3)\times SU(2)\times
U(1)_C\times U(1)_L$\footnote{$U(1)_C={3\over2}U(1)_{B-L};
U(1)_L=2U(1)_{T_{3_R}}.$}
both steps, 1 and 2, are used, in two separate basis vectors.

The breaking of the gauge group and the
reduction to three generations are done simultaneously.
In fact the reduction to three generations is correlated
with the breaking of the flavor $SO(6)^3$ symmetries to a
product of horizontal $U(1)$ symmetries. The appealing
property of the realistic free fermionic models is that
the emergence of three generations is correlated with
the underlying $Z_2\times Z_2$ orbifold structure.
In the superstring standard--like models
each generation is obtained from one of the twisted sectors
of the $Z_2\times Z_2$ orbifold. At the level of the NAHE
set each sector $b_1$, $b_2$ and $b_3$ produces sixteen generations,
transforming as $2\otimes4~\oplus~{\tilde2}\otimes{\bar4}$
under the left and right--moving flavor symmetries
$SO(4)_L\otimes SO(6)_R$. The reduction to three generation
is achieved by breaking the flavor symmetries to $U(1)$ flavor
symmetries and consequently the multiplicity of generations is reduced.
It is clear therefore that it is also possible to construct
models in which the flavor symmetry is not fully broken
and in that case one of the sectors $b_{1,2,3}$ can give more
than one generations while by a suitable choice of GSO projection
coefficient all the generations from the orthogonal sectors
are projected out. Realizing this possibility it is also
possible to construct models with non--Abelian flavor symmetry.

In the models that utilize the full NAHE set and in which
each generation is obtained from one of the sectors
$b_1$, $b_2$ and $b_3$ the flavor symmetries are broken
to $U(1)^n$, with $3\le n\le 9$. Three
$U(1)$ symmetries arise from the complex right--moving
fermions $\bar\eta^1$, $\bar\eta^2$, $\bar\eta^3$. Additional horizontal
$U(1)$ symmetries arise by pairing two of the right--moving real
internal fermions $\{{\bar y},{\bar\omega}\}$. For
every right--moving $U(1)$ symmetry, there is a corresponding
left--moving global $U(1)$ symmetry that is obtained by pairing two of the
left--moving real fermions $\{y,\omega\}$. Each of the remaining
world--sheet left--moving real fermions from the set $\{y,\omega\}$ is paired
with a right--moving real fermion from the set
$\{{\bar y},{\bar\omega}\}$ to form a Ising model operator.
The rank of the final gauge group depends on the number of such
pairings. If all right--moving (and hence left--moving)
fermions were complex, then the gauge group would have rank 22.
Each complexified right--moving pair of real world--sheet fermions
generates a $U(1)$ subgroup of the rank 22 gauge group.
The rank is reduced by the combinations left-- and right--moving
real fermions which form the Ising model operators. We can form
twelve such combinations and therefore the minimal gauge
group has rank sixteen. In addition to fixing the rank of the final
gauge group, this pairing of left-- and right--moving
fermions, through the assignment of boundary conditions,
plays an important role in fixing some of the
low energy properties of the physical spectrum.

To study the construction of the basis vectors beyond the NAHE
set it is convenient to use a notation which emphasizes the
division of the world--sheet fermions by the NAHE set.
This division of the world--sheet fermions is a reflection
of the equivalent underlying $Z_2\times Z_2$ orbifold compactification.
Each one of the sectors $b_1$, $b_2$ and $b_3$
has periodic boundary conditions with respect
to $\{\psi^\mu_{1,2}, \vert {\bar\psi}^{1,\cdots,5}\}$
and one of the sets,
\begin{eqnarray}
\{\chi^{12},y^{3,\cdots,6} \vert{\bar y}^{3,\cdots,6},\bar\eta^1\},\label{s1}\\
\{\chi^{34},y^{1,\cdots,2},{\omega}^{5,6}\vert{\bar y}^{1,2},
			{\bar\omega}^{5,6},{\bar\eta}^2\}\label{s2}\\
\{\chi^{56},{\omega}^{1,\cdots,4}\vert{\bar\omega}^{1,\cdots,4},
						{\bar\eta}^3\}
\label{sets}
\end{eqnarray}
The $\psi_{1,2}^\mu$ are the space--time fermions and
${\bar\psi}^{1,\cdots,5}$ produce the observable $SO(10)$
symmetry. The three complex fermions $\chi^{12}$, $\chi^{34}$
and $\chi^{56}$ correspond to the fermionic superpartners
of the compactified dimensions and carry the space--time
supersymmetry charges. The set of internal fermions
$\{y,\omega\vert{\bar y},{\bar\omega}\}^{1,\cdots,6}$ corresponds
to the left--right symmetric conformal field theory
of the heterotic string, or equivalently to the six--dimensional
compactified manifold in a bosonic formulation.
This set of left/right symmetric internal fermions plays a
fundamental role in the determination of the low energy properties
of the realistic free fermionic models. The analysis of models
beyond the NAHE set is reduced almost entirely to the study
of the boundary conditions of these internal real fermions. Below
I employ a table notation which emphasizes the division of the
internal fermionic states according to their division by the NAHE
set. In the tables, the real fermionic states
$\{y,w\vert{\bar y},{\bar\omega}\}$
are divided according to their division by the NAHE set.
The pairing of real fermions into complex fermions or into
Ising model operators is noted in the table.
The entries in the table represent the boundary conditions
in a basis vector for all the fermionic states.
The basis vectors in a given table are the three
basis vectors which extend the NAHE set.
The notation used is exemplified in table [\ref{modelex}] below.
In the first table the boundary condition of: the space--time
world--sheet fermions, the left--moving complex fermions
$\chi^{12}$, $\chi^{34}$ and $\chi^{56}$ and of the
sixteen complex right--moving world--sheet fermions
${\bar\psi}^{1,\cdots,5},{\bar\eta}^1,{\bar\eta}^2,
{\bar\eta}^3,{\bar\psi}^{1,\cdots,8}$ are shown.
In the second table the boundary conditions under the
world--sheet fermions which correspond to the compactified
space, $\{y,\omega\vert{\bar y},{\bar\omega}\}^{1,\cdots,6}$
are given.
The boundary conditions in the first table therefore
fix the final observable and hidden gauge groups.
The boundary conditions in the second table fix
many of the properties of the low energy observable
spectrum, like for example the number of generations, the rank of the final
gauge group, the presence of Higgs doublets and the projection
of Higgs color triplets, and the non vanishing Yukawa couplings.
The discussion below will therefore mostly focus on the
assignment of boundary conditions to these set of world--sheet
fermions and the boundary conditions of the remaining world--sheet
fermions will sometimes not be shown explicitly.

\beqn
 &\begin{tabular}{c|c|ccc|c|ccc|c}
 ~ & $\psi^\mu$ & $\chi^{12}$ & $\chi^{34}$ & $\chi^{56}$ &
        $\bar{\psi}^{1,...,5} $ &
        $\bar{\eta}^1 $&
        $\bar{\eta}^2 $&
        $\bar{\eta}^3 $&
        $\bar{\phi}^{1,...,8} $ \\
\hline
\hline
  ${b_4}$     &  1 & 1&0&0 & 1~1~1~1~1 & 1 & 0 & 0 & 0~0~0~0~0~0~0~0 \\
  ${b_5}$     &  1 & 0&1&0 & 1~1~1~1~1 & 0 & 1 & 0 & 0~0~0~0~0~0~0~0 \\
  ${\gamma}$  &  1 & 0&0&1 &
		${1\over2}$~${1\over2}$~${1\over2}$~${1\over2}$~${1\over2}$
		& ${1\over2}$ & ${1\over2}$ & ${1\over2}$ &
                ${1\over2}$~${1\over2}$~${1\over2}$~${1\over2}$~1~0~0~0 \\
\end{tabular}
   \nonumber\\
   ~  &  ~ \nonumber\\
   ~  &  ~ \nonumber\\
     &\begin{tabular}{c|c|c|c}
 ~&   $y^3{\bar y}^3$
      $y^4{\bar y}^4$
      $y^5{\bar y}^5$
      $y^6{\bar y}^6$
  &   $y^1{\bar y}^1$
      $y^2{\bar y}^2$
      $\omega^5{\bar\omega}^5$
      $\omega^6{\bar\omega}^6$
  &   $\omega^1{\bar\omega}^1$
      $\omega^2{\bar\omega}^2$
      $\omega^3{\bar\omega}^3$
      $\omega^4{\bar\omega}^4$ \\
\hline
\hline
$b_4$  & 1 ~~~ 0 ~~~ 0 ~~~ 1  & 0 ~~~ 0 ~~~ 1 ~~~ 0  & 0 ~~~ 0 ~~~ 0 ~~~ 1 \\
$b_5$  & 0 ~~~ 0 ~~~ 0 ~~~ 1  & 0 ~~~ 1 ~~~ 1 ~~~ 0  & 1 ~~~ 0 ~~~ 0 ~~~ 0 \\
$\gamma$ & 1 ~~~ 1 ~~~ 0 ~~~ 0 & 1 ~~~ 0 ~~~ 0 ~~~ 0 & 0 ~~~ 1 ~~~ 0 ~~~ 0 \\
\end{tabular}
\label{modelex}
\eeqn
Note that in the notation used here, basis vectors which preserve
the $SO(10)$ symmetry are denoted by $b_j$ $(j=4,5,...)$
while basis vectors which break the $SO(10)$ symmetry are denoted
by small Greek letters. The model of table [\ref{modelex}] is
an example of a three generation $SU(5)\times U(1)$ model. The
twelve real left--moving fermions $\{y,\omega\}$ are combined
with twelve real right--moving fermions
$\{{\bar y},{\bar\omega}\}$ to form twelve
Ising model operators. Therefore, the rank of the final
right--moving gauge group in this model is sixteen.

The boundary condition basis vectors span a finite additive group. Then
any subset of the independent vectors of the additive group
will span the same model, up to the choice of GSO projection
coefficients. Given that the additive group contains
typically $2^7\times4$ sectors the question then is how
can we avoid reproducing the same models. In computerized searches
this problem is addressed in \cite{senecal}. However, when trying to
develop an insight how the boundary conditions
fix the physical properties of the models a computerized
search is not suitable. The problem is avoided by requiring
that the basis vectors of a new model cannot all be obtained
by a linear combination of the basis vectors of a previous model.
Each new model must contain at least one basis vector
which cannot be realized as a linear combination of the basis
vectors of a previous model. The range of allowed basis
vectors is still however very large. In addition each choice
of basis vectors can span a set of distinct models by the
discrete choices of free phases, as will be illustrated below.

\setcounter{footnote}{0}
\section{Classification by boundary condition rules}
In this section I discuss the role of the boundary conditions
in the determination of the low energy properties of the superstring models.
The purpose here is to illustrate how several of the
properties of the low energy
spectrum are determined by general boundary condition rules.
As some of these results have already appeared previously in the
literature, the discussion will be concise.

\subsection{Higgs doublet--triplet splitting}\label{higgs}

The Higgs doublet--triplet splitting operates as follows.
The Neveu--Schwarz sector gives rise to three fields in the
10 representation of $SO(10)$.  These contain the  Higgs electroweak
doublets and color triplets. Each of those is charged with respect to one
of the horizontal $U(1)$ symmetries $U(1)_{1,2,3}$.  Each one of these
multiplets is associated, by the horizontal symmetries, with one of the
twisted sectors, $b_1$, $b_2$ and $b_3$. The doublet--triplet
splitting results from the boundary condition basis vectors which breaks
the $SO(10)$ symmetry to $SO(6)\times SO(4)$. We can define a quantity
$\Delta_i$ in these basis vectors which measures the difference between the
boundary conditions assigned to the internal fermions from the set
$\{y,w\vert{\bar y},{\bar\omega}\}$ and which are periodic in the vector
$b_i$,
\begin{equation}
\Delta_i=\vert\alpha_L({\rm internal})-
\alpha_R({\rm internal})\vert=0,1~~(i=1,2,3)
\label{dts}
\end{equation}
If $\Delta_i=0$ then the Higgs triplets, $D_i$ and ${\bar D}_i$,
remain in the massless spectrum while the Higgs doublets, $h_i$ and ${\bar
h}_i$ are projected out
and the opposite occurs for $\Delta_i=1$.

Thus, the rule in Eq. (\ref{dts})
is a generic rule that can be used in the construction
of the free fermionic models, with the NAHE set. The model
of table [\ref{colorhiggs}] illustrates this rule.
In this model $\Delta_1=\Delta_2=0$ while $\Delta_3=1$. Therefore,
this model produces two pairs of color triplets and one pair of
Higgs doublets from the Neveu--Schwarz sector, $D_1$, $\bar D_1$
$D_2$, $\bar D_2$ and $h_3$, $\bar h_3$.
\beqn
 &\begin{tabular}{c|c|ccc|c|ccc|c}
 ~ & $\psi^\mu$ & $\chi^{12}$ & $\chi^{34}$ & $\chi^{56}$ &
        $\bar{\psi}^{1,...,5} $ &
        $\bar{\eta}^1 $&
        $\bar{\eta}^2 $&
        $\bar{\eta}^3 $&
        $\bar{\phi}^{1,...,8} $ \\
\hline
\hline
  ${\alpha}$  &  1 & 1&0&0 & 1~1~1~0~0 & 1 & 0 & 1 & 1~1~1~1~0~0~0~0 \\
  ${\beta}$   &  1 & 0&1&0 & 1~1~1~0~0 & 0 & 1 & 1 & 1~1~1~1~0~0~0~0 \\
  ${\gamma}$  &  1 & 0&0&1 &
		${1\over2}$~${1\over2}$~${1\over2}$~${1\over2}$~${1\over2}$
	      & ${1\over2}$ & ${1\over2}$ & ${1\over2}$ &
                ${1\over2}$~0~1~1~${1\over2}$~${1\over2}$~${1\over2}$~0 \\
\end{tabular}
   \nonumber\\
   ~  &  ~ \nonumber\\
   ~  &  ~ \nonumber\\
     &\begin{tabular}{c|c|c|c}
 ~&   $y^3{\bar y}^3$
      $y^4{\bar y}^4$
      $y^5{\bar y}^5$
      $y^6{\bar y}^6$
  &   $y^1{\bar y}^1$
      $y^2{\bar y}^2$
      $\omega^5{\bar\omega}^5$
      $\omega^6{\bar\omega}^6$
  &   $\omega^2{\omega}^3$
      $\omega^1{\bar\omega}^1$
      $\omega^4{\bar\omega}^4$
      ${\bar\omega}^2{\bar\omega}^3$ \\
\hline
\hline
$\alpha$ & 1 ~~~ 0 ~~~ 0 ~~~ 1  & 0 ~~~ 0 ~~~ 1 ~~~ 0  & 0 ~~~ 0 ~~~ 1 ~~~ 1 \\
$\beta$  & 0 ~~~ 0 ~~~ 0 ~~~ 1  & 0 ~~~ 1 ~~~ 1 ~~~ 0  & 0 ~~~ 1 ~~~ 0 ~~~ 1 \\
$\gamma$ & 1 ~~~ 1 ~~~ 0 ~~~ 0  & 1 ~~~ 1 ~~~ 0 ~~~ 0  & 0 ~~~ 0 ~~~ 0 ~~~ 1 \\
\end{tabular}
\label{colorhiggs}
\eeqn
With the choice of generalized GSO coefficients:
\beqn
&& c\left(\matrix{b_1,b_3,\alpha,\beta,\gamma\cr
                                    \alpha\cr}\right)=
-c\left(\matrix{b_2\cr
                                    \alpha\cr}\right)=
c\left(\matrix{{\bf1},b_j,\gamma\cr
                                    \beta\cr}\right)=\nonumber\\
&& c\left(\matrix{\gamma\cr
                                   b_3\cr}\right)=
-c\left(\matrix{\gamma\cr
                                    {\bf1},b_1,b_2\cr}\right)=-1\nonumber
\eeqn
(j=1,2,3), with the others specified by modular invariance and space--time
supersymmetry.

Another relevant question with regard to the Higgs doublet--triplet
splitting mechanism is whether it is possible to construct models in which
both the Higgs color triplets and electroweak doublets from the
Neveu--Schwarz sector are projected out by the GSO projections.
This is a viable possibility as we can choose for example
$$\Delta_j^{(\alpha)}=1 ~{\rm and}~ \Delta_j^{(\beta)}=0,$$
where $\Delta^{(\alpha,\beta)}$ are the projections due
to the basis vectors $\alpha$ and $\beta$ respectively.
This is a relevant question as the number of Higgs representations,
which generically appear in the massless spectrum,
is larger than what is allowed by the low energy phenomenology.
Consider for example the model in table [\ref{bothdandt}]
\beqn
 &\begin{tabular}{c|c|ccc|c|ccc|c}
 ~ & $\psi^\mu$ & $\chi^{12}$ & $\chi^{34}$ & $\chi^{56}$ &
        $\bar{\psi}^{1,...,5} $ &
        $\bar{\eta}^1 $&
        $\bar{\eta}^2 $&
        $\bar{\eta}^3 $&
        $\bar{\phi}^{1,...,8} $ \\
\hline
\hline
  ${\alpha}$  &  1 & 1&0&0 & 1~1~1~0~0 & 1 & 0 & 1 & 1~1~1~1~0~0~0~0 \\
  ${\beta}$   &  1 & 0&1&0 & 1~1~1~0~0 & 0 & 1 & 1 & 1~1~1~1~0~0~0~0 \\
  ${\gamma}$  &  1 & 0&0&1 &
		${1\over2}$~${1\over2}$~${1\over2}$~${1\over2}$~${1\over2}$
	      & ${1\over2}$ & ${1\over2}$ & ${1\over2}$ &
                ${1\over2}$~0~1~1~${1\over2}$~${1\over2}$~${1\over2}$~0 \\
\end{tabular}
   \nonumber\\
   ~  &  ~ \nonumber\\
   ~  &  ~ \nonumber\\
     &\begin{tabular}{c|c|c|c}
 ~&   $y^3{y}^6$
      $y^4{\bar y}^4$
      $y^5{\bar y}^5$
      ${\bar y}^3{\bar y}^6$
  &   $y^1{\omega}^5$
      $y^2{\bar y}^2$
      $\omega^6{\bar\omega}^6$
      ${\bar y}^1{\bar\omega}^5$
  &   $\omega^2{\omega}^4$
      $\omega^1{\bar\omega}^1$
      $\omega^3{\bar\omega}^3$
      ${\bar\omega}^2{\bar\omega}^4$ \\
\hline
\hline
$\alpha$ & 1 ~~~ 0 ~~~ 0 ~~~ 0  & 0 ~~~ 0 ~~~ 1 ~~~ 1  & 0 ~~~ 0 ~~~ 1 ~~~ 1 \\
$\beta$  & 0 ~~~ 0 ~~~ 1 ~~~ 0  & 1 ~~~ 0 ~~~ 0 ~~~ 0  & 0 ~~~ 1 ~~~ 0 ~~~ 0 \\
$\gamma$ & 0 ~~~ 1 ~~~ 0 ~~~ 0  & 0 ~~~ 1 ~~~ 0 ~~~ 1  & 1 ~~~ 0 ~~~ 0 ~~~ 0 \\
\end{tabular}
\label{bothdandt}
\eeqn
With the choice of generalized GSO coefficients:
\beqn
&& c\left(\matrix{b_j\cr
                           S,b_j,\alpha,\beta,\gamma\cr}\right)=
c\left(\matrix{\alpha\cr
                           \alpha,\beta,\gamma\cr}\right)=\nonumber\\
&&c\left(\matrix{\beta,\cr
                                    \beta,\gamma\cr}\right)=
 c\left(\matrix{\gamma\cr
                                   {\bf1}\cr}\right)=-1\nonumber
\eeqn
(j=1,2,3), with the others specified by modular invariance and space--time
supersymmetry. In this model
$\Delta_1^{(\alpha)}=\Delta_2^{(\alpha)}=\Delta_3^{(\alpha)}=1$,
and $\Delta_1^{(\beta)}=\Delta_3^{(\beta)}=0$,
Therefore, In this model irrespective of the choice of the generalized GSO
projection coefficients, both the Higgs color triplets and electroweak
doublets associated with $b_1$ and $b_3$ are projected out by the
GSO projections. However, it is found that the combination of these
projections also results in the projection of some
of the representations from the corresponding sectors
$b_1$ and $b_3$ and therefore these sectors do not produce
the full chiral 16 of $SO(10)$. Therefore, realization of
this mutual projection of both Higgs triplets and doublets
from the Neveu--Schwarz sector requires that the chiral
generations be obtained from non--NAHE set basis vectors.

\subsection{Cubic level Yukawa couplings}\label{yukcop}
As a second example I discuss how the boundary condition
basis vectors fix the cubic level Yukawa couplings for the quarks
and leptons. These Yukawa couplings are fixed by the vector $\gamma$
which breaks the $SO(10)$ symmetry to $SU(5)\times U(1)$.  Each sector
$b_i$ gives rise to an up--like or down--like cubic level Yukawa coupling.
We can define a similar quantity $\Delta$ in the vector $\gamma$,
which breaks the $SO(10)$ symmetry to $SU(5)\times U(1)$.
The quantity $\Delta$ again measures the difference between
the left-- and right--moving
boundary conditions assigned to the internal fermions from the set
$\{y,w\vert{\bar y},{\bar\omega}\}$ and which are periodic in the vector
$b_i$,
\begin{equation}
\Delta_i=\vert\gamma_L({\rm internal})-
\gamma_R({\rm internal})\vert=0,1~~(i=1,2,3)
\label{udyc}
\end{equation}
If $\Delta_i=0$ then the sector $b_i$ gives rise to a
down--like Yukawa coupling while the
up--type Yukawa coupling vanishes. The opposite occurs if $\Delta_i=1$.
The model of table [\ref{yukawa}] illustrates how this rule
works. In this model $\Delta_1=\Delta_3=1$ and $\Delta_2=0$.
Therefore in this model the sectors $b_1$ and $b_3$ produce
up type quark Yukawa couplings while the sector $b_2$ produces Yukawa
couplings for the down type quark and for the charged lepton.
\beqn
 &\begin{tabular}{c|c|ccc|c|ccc|c}
 ~ & $\psi^\mu$ & $\chi^{12}$ & $\chi^{34}$ & $\chi^{56}$ &
        $\bar{\psi}^{1,...,5} $ &
        $\bar{\eta}^1 $&
        $\bar{\eta}^2 $&
        $\bar{\eta}^3 $&
        $\bar{\phi}^{1,...,8} $ \\
\hline
\hline
  ${\alpha}$  &  0 & 0&0&0 & 1~1~1~0~0 & 0 & 0 & 0 & 1~1~1~1~0~0~0~0 \\
  ${\beta}$   &  0 & 0&0&0 & 1~1~1~0~0 & 0 & 0 & 0 & 1~1~1~1~0~0~0~0 \\
  ${\gamma}$  &  0 & 0&0&0 &
		${1\over2}$~${1\over2}$~${1\over2}$~${1\over2}$~${1\over2}$
	      & ${1\over2}$ & ${1\over2}$ & ${1\over2}$ &
                ${1\over2}$~0~1~1~${1\over2}$~${1\over2}$~${1\over2}$~1 \\
\end{tabular}
   \nonumber\\
   ~  &  ~ \nonumber\\
   ~  &  ~ \nonumber\\
     &\begin{tabular}{c|c|c|c}
 ~&   $y^3{y}^6$
      $y^4{\bar y}^4$
      $y^5{\bar y}^5$
      ${\bar y}^3{\bar y}^6$
  &   $y^1{\omega}^5$
      $y^2{\bar y}^2$
      $\omega^6{\bar\omega}^6$
      ${\bar y}^1{\bar\omega}^5$
  &   $\omega^2{\omega}^4$
      $\omega^1{\bar\omega}^1$
      $\omega^3{\bar\omega}^3$
      ${\bar\omega}^2{\bar\omega}^4$ \\
\hline
\hline
$\alpha$ & 1 ~~~ 1 ~~~ 1 ~~~ 0  & 1 ~~~ 1 ~~~ 1 ~~~ 0  & 1 ~~~ 1 ~~~ 1 ~~~ 0 \\
$\beta$  & 1 ~~~ 0 ~~~ 0 ~~~ 0  & 0 ~~~ 0 ~~~ 1 ~~~ 1  & 0 ~~~ 0 ~~~ 1 ~~~ 1 \\
$\gamma$ & 0 ~~~ 0 ~~~ 1 ~~~ 1  & 1 ~~~ 0 ~~~ 0 ~~~ 1  & 0 ~~~ 1 ~~~ 0 ~~~ 1 \\
\end{tabular}
\label{yukawa}
\eeqn
With the choice of generalized GSO coefficients:
\beqn
&& c\left(\matrix{b_1,b_3\cr
                                    \alpha,\beta,\gamma\cr}\right)=
-c\left(\matrix{\gamma\cr
                                    b_2\cr}\right)=
-c\left(\matrix{\alpha\cr
                                    \alpha,\beta\cr}\right)=\nonumber\\
&& c\left(\matrix{\gamma\cr
                                    \gamma\cr}\right)=
c\left(\matrix{\beta\cr
                                    \beta\cr}\right)=
-c\left(\matrix{\gamma\cr
                                    \alpha,\beta\cr}\right)=-1\nonumber
\eeqn
$(j=1,2,3),$
with the others specified by modular invariance and space--time
supersymmetry. In this model therefore we obtain at the cubic level
$Q_1u_1{\bar h}_1$, $L_1N_1{\bar h}_1$,
$Q_2d_2{h}_2$, $L_2e_2{h}_2$ and
$Q_3u_3{\bar h}_3$, $L_3N_3{\bar h}_3$,
irrespective of the choice of  GSO projection coefficients.

Another point that should be noted is that there
is a $Z_2$ ambiguity in the definition of the weak
hypercharge. We can define the weak hypercharge by
\beq
U(1)_Y={1\over3}U(1)_C\pm {1\over2}U(1)_L
\label{weakhyper}
\eeq
where the $+$ choice is the one typically chosen in the
literature. The alternative choice corresponds to
the flipping of the representations
\beqn
&& e_L^c ~\leftrightarrow~ N_L^c\nonumber\\
&& u_L^c ~\leftrightarrow~ d_L^c\nonumber
\label{repflip}
\eeqn
This flip is equivalent to the flip between the straight
and flipped $SU(5)$ representations. In the case of $SU(5)$
only the later choice is allowed as there are no adjoint representations
to break the non--Abelian gauge symmetry in the former.
In the case of
the standard--like models, as the GUT non--Abelian symmetry
is broken directly at the string level, this flip can
be consistent with the low energy constraints.
We note however that under this $Z_2$ flip the Higgs representations
are also flipped $h\leftrightarrow{\bar h}$ and therefore
the Yukawa coupling rule is invariant under this ambiguity
in the definition of the weak hypercharge.

The above two examples illustrate how general rules can
be extracted which determine how the boundary conditions
fix the low energy phenomenological properties of the
string models. These results
can be proven by using the definition of the GSO projection
and the modular invariance rules \cite{yukawa,ps}.

\setcounter{footnote}{0}
\section{Classification by the role of the free phases}

The focus in this section will be on the effect of the
one--loop GSO phases
$$c\left({\matrix{a_i\cr a_j\cr}}\right)$$
on the properties of the
massless spectrum of the realistic free fermionic models.
Here $a_i$, $a_j$ represent the NAHE set, and the additional,
boundary conditions basis vectors which generate the realistic
free fermionic models.

\subsection{General remarks}
Like with the boundary conditions it is convenient to
split the free phases into the NAHE and
beyond the NAHE sectors. In Eq. (\ref{freephases})
I employ a matrix notation for the free phases
between all the boundary condition basis vectors. For example, in the model
of ref. \cite{fny} the following choice of free phases has been made,
\begin{equation}
{\bordermatrix{
              &  1&  S & & {b_1}&{b_2}&{b_3}& & {\alpha}&{\beta}&{\gamma}\cr
             1& -1&~~1 & & -1   &  -1 & -1  & & ~~1     &  -1   & ~~i   \cr
             S&~~1&~~1 & &~~1   & ~~1 &~~1  & & ~~1     & ~~1   & ~~1   \cr
	      &   &    & &      &     &     & &         &       &       \cr
         {b_1}& -1& -1 & & -1   &  -1 & -1  & &  -1     &  -1   & ~~i   \cr
         {b_2}& -1& -1 & & -1   &  -1 & -1  & &  -1     &  -1   & ~~i   \cr
         {b_3}& -1& -1 & & -1   &  -1 & -1  & &  -1     &  -1   &  -1   \cr
	      &   &    & &      &     &     & &         &       &       \cr
      {\alpha}&~~1& -1 & & -1   &  -1 & -1  & & ~~1     &  -1   & ~~i   \cr
       {\beta}& -1& -1 & & -1   &  -1 & -1  & &  -1     &  -1   & ~~1   \cr
            			                		         {\gamma}      &~~1& -1 & &~~1   &  -1
&~~1  & & ~~1     &  -1   & ~~1 \cr}}
\label{freephases}
\end{equation}
Only the entries above the diagonal are
independent and those below and on the diagonal are fixed by
the modular invariance constraints. In the matrix above
blank lines are inserted to emphasize the division of the free
phases between the different sectors of the realistic
free fermionic models. Thus, the first two lines involve
only the GSO phases of $c{{\{{\bf 1},S\}}\choose a_i}$. The set
$\{{\bf 1},S\}$ generates the $N=4$ model with $S$ being the
space--time supersymmetry generator and therefore the phases
$c{S\choose{a_i}}$ are those that control the space--time supersymmetry
in the superstring models. Similarly, in the free fermionic
models, sectors with periodic and anti--periodic boundary conditions,
of the form of $b_i$, produce the chiral generations.
The phases $c{b_i\choose b_j}$ determine
the chirality of the states from these sectors, and as will be shown
below give rise to the mirror symmetry. Likewise, in the free fermionic
models the basis vector $b_i$ are those that respect the $SO(10)$ symmetry
while the vectors denoted by Greek letters are those that break the
$SO(10)$ symmetry. The phenomenology of the Standard Model
sector including the texture of fermion mass matrices
is obtained exclusively from the untwisted sector, the sectors $b_i$
and $b_i+2\gamma$ and the sector $b_1+b_2+\alpha+\beta$. These are
the sectors which preserve the $SO(10)$ symmetry. Thus,
the phases that play a role in the phenomenology of the Standard Model
sector are, in general, the phases $c{b_i\choose{a_i}}$. On the other
hand the basis vectors of the form $\{\alpha,\beta,\gamma\}$ break the
$SO(10)$ symmetry. The phases associated with these basis vectors
in general are associated with the exotic physics sectors
that go beyond the Standard Model. These sectors for example
control the number of additional exotic color triplets, beyond
the spectrum of the Minimal Supersymmetric Standard Model. Therefore the
phases associated with these basis vectors will be important, for example,
with regard to the problem of string scale gauge coupling unification.

\subsection{$E_6~\rightarrow~SO(10)$ breaking}\label{twozerosec}

The NAHE set plus the vector $X$ produce a model with
$SO(4)^3\times E_6\times U(1)^2\times E_8$ gauge group and with (2,2)
world--sheet supersymmetry. The realistic free fermionic models have
at the level of the NAHE set an $SO(10)$ symmetry and only $(2,0)$
world--sheet supersymmetry. However, the realistic free fermionic
models have some underlying (2,2)
structure which is why we can identify the internal fermions
$\{y, \omega\vert{\bar y},{\bar\omega}\}$ with the compactified
dimensions in the bosonic
formulation. In the realistic free fermionic models the basis
vector $2\gamma$ replaces the vector $X$.
The set $\{{\bf 1},S,I={\bf 1}+b_1+b_2+b_3,2\gamma\}$
produces a model with $N=4$ supersymmetry and $SO(12)\times
SO(16)\times SO(16)$ gauge group. Alternatively, we can construct
the same model by using the set $\{{\bf 1},S,X,I\}$. Using this set
we can construct two $N=4$ models which differ by the discrete
choice of the free phase
\begin{equation}
c\left(\matrix{X\cr
                        I\cr}\right)=\pm1
\label{22to20}
\end{equation}
For one choice of this phase the gauge group is $SO(12)\times
E_8\times E_8$ which is the standard toroidal compactification.
However, for the second choice of this phase the states in the
spinorial representation of $SO(16)$ which make up the adjoint
of $E_8$ are projected out by the GSO projections. Thus,
we are left with $SO(12)\times SO(16)\times SO(16)$ gauge
group.  Applying the orbifold twisting on the first choice,
with $E_8\times E_8$ gauge group, and with standard embedding,
breaks one of the $E_8$ to $E_6\times U(1)^2$ and gives
rise to the models with $(2,2)$ world--sheet supersymmetry.
Applying the same orbifold twisting to the second discrete choice
results in the models with $SO(10)\times U(1)^3$ gauge group.
Alternatively, we can start with the basis vectors that
produce the $E_6\times U(1)^2$
model and then turn on the GSO projection which projects out
the spinorial $16+{\overline{16}}$ in the adjoint of $E_6$.
We end with the same model, as the spectrum is invariant
under the reordering of the GSO projections.
Thus, the breaking of the right--moving $N=2$ world--sheet
supersymmetry can be seen to be a result of a discrete choice
of the free phases. Further, analysis of this breaking
in connection with the anomalous $U(1)_A$ which appears
in these models, will be reported in ref. \cite{cf}.
This is an important observation because many
of the useful simple relations that are obtained for $(2,2)$ models
can be used for the realistic free fermionic models. It is precisely for
this reason that the set of real fermions
$\{y,\omega\vert{\bar y},{\bar\omega}\}$
can be identified with the six left-- and
right--moving compactified dimensions.

\subsection{free phases and mirror symmetry}\label{mirrorsec}
Mirror symmetry in the free fermionic models is a result of the choices
of free phases. To see how this works I consider for example
the (2,2) model generated by the set
$$\{{\bf 1},S, X,I,b_1,b_2\}.$$
with the choice of free phases
\begin{equation}
c\left(\matrix{b_i\cr
                                    b_j\cr}\right)=
c\left(\matrix{b_i\cr
                                    S\cr}\right)=
c\left(\matrix{b_i\cr
                                    I,X\cr}\right)=
-c\left(\matrix{{\bf1}\cr
                                    I,X\cr}\right)=
-c\left(\matrix{I\cr
                                    X\cr}\right)=
-c\left(\matrix{{\bf1}\cr
                                    {\bf1}\cr}\right)=-1,
\label{e6model}
\end{equation}
In this model the gauge group is $SO(4)^3\times
E_6\times U(1)^2\times E_8$ with $N=1$ space--time supersymmetry.
There are 24 generations in the 27 representation of
$E_6$, eight from each twisted sector. In the fermionic construction
these are the sectors $(b_1;b_1+X)$, $(b_2;b_2+X)$ and $(b_3;b_3+X)$,
where the sectors $b_j$ produce the 16 and the sectors $b_j+X$ produce
the $10\oplus1$ representations in the decomposition of the $E_6$
representations under $SO(10)\times U(1)$.

In this model the only internal fermionic states which count the
multiplets of $E_6$ are the real internal fermions $\{y,w\vert{\bar y},
{\bar\omega}\}$. This is observed by writing the degenerate vacuum
of the sectors $b_j$ in a combinatorial notation. The vacuum of the sectors
$b_j$  contains twelve periodic fermions. Each periodic fermion
gives rise to a two dimensional degenerate vacuum $\vert{+}\rangle$ and
$\vert{-}\rangle$ with fermion numbers $0$ and $-1$, respectively.
The GSO operator, is a generalized parity, operator which
selects states with definite parity. After applying the
GSO projections, we can write the degenerate vacuum of the sector
$b_1$ in combinatorial form
\begin{eqnarray}
{\left[\left(\matrix{4\cr
                                    0\cr}\right)+
\left(\matrix{4\cr
                                    2\cr}\right)+
\left(\matrix{4\cr
                                    4\cr}\right)\right]
\left\{\left(\matrix{2\cr
                                    0\cr}\right)\right.}
&{\left[\left(\matrix{5\cr
                                    0\cr}\right)+
\left(\matrix{5\cr
                                    2\cr}\right)+
\left(\matrix{5\cr
                                    4\cr}\right)\right]
\left(\matrix{1\cr
                                    0\cr}\right)}\nonumber\\
{+\left(\matrix{2\cr
                                    2\cr}\right)}
&{~\left[\left(\matrix{5\cr
                                    1\cr}\right)+
\left(\matrix{5\cr
                                    3\cr}\right)+
\left(\matrix{5\cr
                                    5\cr}\right)\right]\left.
\left(\matrix{1\cr
                                    1\cr}\right)\right\}}
\label{spinor}
\end{eqnarray}
where
$4=\{y^3y^4,y^5y^6,{\bar y}^3{\bar y}^4,
{\bar y}^5{\bar y}^6\}$, $2=\{\psi^\mu,\chi^{12}\}$,
$5=\{{\bar\psi}^{1,\cdots,5}\}$ and $1=\{{\bar\eta}^1\}$.
The combinatorial factor counts the number of $\vert{-}\rangle$ in the
degenerate vacuum of a given state.
The two terms in the curly brackets correspond to the two
components of a Weyl spinor.  The $10+1$ in the $27$ of $E_6$ are
obtained from the sector $b_j+X$.
{}From Eq. (\ref{spinor}) it is observed that the states
which count the multiplicities of $E_6$ are the internal
fermionic states $\{y^{3,\cdots,6}\vert{\bar y}^{3,\cdots,6}\}$.
A similar result is
obtained for the sectors $b_2$ and $b_3$ with $\{y^{1,2},\omega^{5,6}
\vert{\bar y}^{1,2},{\bar\omega}^{5,6}\}$
and $\{\omega^{1,\cdots,4}\vert{\bar\omega}^{1,\cdots,4}\}$
respectively, which indicates that
these twelve states correspond to a six dimensional
compactified orbifold with Euler characteristic equal to 48.

The chirality of the states from a twisted sector $b_j$
is fixed by the GSO projection induced by another twisted
sector $b_i$, with $i\ne j$. The chirality under
the space--time spinor group is determined by the
sign of the vacuum of the space--time fermion $\psi^\mu$.
To obtain the chiral representation of the $E_6$ gauge group,
{\it i.e.} to project the ${\overline{27}}$ from the spectrum a
``chirality condition'' is obeyed \cite{nahe}. This ``chirality condition''
states that in the sector $b_i$, which induces the chirality projection,
the world--sheet fermions $\{\psi^\mu,{\bar\psi}^{1,\cdots,5}\}$
are periodic. The remaining world--sheet fermions which are
periodic in the vector $b_j$ are antiperiodic in the vector
$b_i$. Thus, the chirality of the states from a twisted sector
$b_j$ is determined by the free phase $c{b_j\choose b_i}$.
Since we have a freedom in the choice of the sign of this
free phase, we can get from the sector $(b_j;b_j+X)$ either the
27 or the $\overline{27}$. Which of those we obtain in the physical
spectrum depends on the sign of the free phase. Thus, mirror symmetry
in the free fermionic construction is the statement
\begin{equation}
c{b_j\choose b_i}~\rightarrow~-c{b_j\choose b_i}
\label{mirror}
\end{equation}
The free phases $c{b_j\choose b_i}$ also fix the total number
of chiral generations. Since there are two $b_i$ projections
for each sector $b_j$, $i\ne j$ we can use one projections to project
out the states with one chirality and the other projection to
project out the states with the other chirality. Thus, the total
number of generations with this set of basis vectors
is given by
$$8({{c{b_1\choose b_2}+c{b_1\choose b_3}}\over2})+
 8({{c{b_2\choose b_1}+c{b_2\choose b_3}}\over2})+
 8({{c{b_3\choose b_1}+c{b_3\choose b_2}}\over2})$$
Since the modular invariance rules fix
$c{b_j\choose b_i}=c{b_i\choose b_j}$ we get that
the total number of generations is either
24 or 8.

We can now see that the same mirror phenomena
also holds in the case of the $(2,0)$ models.
As the first example consider the case in which
we start with the $(2,2)$ model and then
project the $16\oplus{\overline{16}}$ generators in the
adjoint of $E_6$. As explained above,
this is achieved by the choice
of the free phase or by replacing the vector $X$
with the vector $2\gamma$. The gauge group now is
$SO(10)\times U(1)^3\times \times SO(4)^3\times SO(16)$
In this case the $10\oplus1$ states
from the sectors $b_j+X$ are projected out from the massless spectrum
and instead we obtain the states in the vectorial $16$ representations
of the hidden $SO(16)$ gauge group. The sectors $b_j$ still produce
the chiral 16 representation of $SO(10)$, with the same vacuum
degeneracy as in Eq. (\ref{spinor}). The chirality is obtained
in the same way and therefore the mirror models are obtained
by taking $c{b_j\choose b_i}~\rightarrow~-c{b_j\choose b_i}$
as in the (2,2) model. Thus, in this case the (2,2) structure
is still preserved after the projection. The world--sheet
fermions $\{y,\omega\vert{\bar y},{\bar\omega}\}$
are identified with the compactified dimensions.

The mirror symmetry as a consequence of the free phases
is also observed in models that do not preserve the $(2,2)$
structure. Consider for example the model generated by the NAHE
set, $\{{\bf 1},S,b_1,b_2,b_3\}$. This model has a gauge group
$SO(10)\times SO(6)^3\times E_8$ gauge symmetry. The world--sheet
fermions $\{{\bar y},{\bar\omega}\}$ and ${\bar\eta}^{1,2,3}$
are mixed and produce the horizontal $SO(6)^3$ symmetries.
However, also in this case the same ``chirality condition''
operates and the chiral spectrum is produced by the projection
of $b_j$ on $b_i$, $i\ne j$. In this case we obtain twice the
number of generations, namely 48 (16 from each sector $b_j$).
In addition to the states in Eq.(\ref{spinor}) we
have the states with the spin $-1/2$ vacuum for ${\bar\eta}^j$
and with parity $-1$ for the fermions from the set
$\{y,\omega\vert{\bar y},{\bar\omega}\}^{1,\cdots,6}$.
In this model as well the mirror families are obtained
by the transformation $c{b_j\choose b_i}\rightarrow -c{b_j\choose b_i}$.
Thus, depending on the choice of the free phases we can
construct with this choice of boundary condition basis vectors,
models with 48 or 16 generations and their mirrors.
The mirror symmetry operates in this model in the same way
that it does in the $(2,2)$ model and in the previous $(2,0)$ model
which preserves the $(2,2)$ structure.

The mirror symmetry in the free fermionic models with
$Z_2$ twists is therefore seen to be a result of the freedom to choose
the free GSO phases. In general, for any sector of the form
of the sectors $b_j$ we can get the opposite chirality by
flipping the sign of the phase $c{b_j\choose b_i}$
provided that the sectors $b_j$ and $b_i$ satisfy the ``chirality
condition''. Provided that all the vectors which satisfy the
``chirality condition'' with the sector $b_j$ have the same
sign for $c{b_j\choose b_i}$, insures that the states from the
sector $b_j$ are not completely projected out. The mirror model
of a given model is obtained by taking
$c{b_i\choose b_j}~\rightarrow~-c{b_i\choose b_j}$ for
all $i$ and $j$.

It is important to note that the mirror symmetry in terms of
the free phases corresponds to the same phenomena in terms of
discrete torsion in the orbifold models \cite{vafa,vafawitten}.
However, we note that the description in terms of the free fermions
seems to be richer than the corresponding orbifold
construction. In the orbifold model we can get a model
and its mirror by turning on the discrete torsion.
However, in the fermionic models, in addition
to a model and its mirror obtained by a global
GSO phase change
$c{b_j\choose b_i}~\rightarrow~-c{b_i\choose b_j}$ for
all $i$ and $j$, we have the models which are not related by
the change of the sign of the Euler characteristic, but in which
the Euler characteristic is modified discretely.
It is intriguing also to note that similar ambiguity
in terms of discrete--like torsion may also play a role
in string strong--weak coupling duality \cite{MandFcompact}.

\subsection{Free phases and space--time supersymmetry}

In this subsection I briefly discuss the relation between
free phases and space--time supersymmetry.
While supersymmetry is not yet an observed symmetry of nature
it does have many attractive features which motivates the
search for superparticles at the electroweak scale.
Supersymmetry for example explains the origin of the
electroweak symmetry breaking scale in terms of the
higher unification scale and the evolution of the
couplings to the low scale. This is an important property
and the recent observation of the top quark mass
is in good agreement with the expectations from supersymmetry.
Another important aspect of supersymmetry is that it naturally
arises in superstring theory which at present is the leading
candidate for a theory of quantum gravity. However
as the breaking of supersymmetry, in this context,
is still obscured, it is important to search for
alternative models where supersymmetry is broken at a high scale.
For example in ref. \cite{dienes} this option has been explored and
it has been shown that even if supersymmetry is broken near
the Planck scale there still exist a ``misaligned supersymmetry''
that mixes between states across mass levels.

In the realistic free fermionic models the space--time supersymmetry
is obtained from the basis vector $S$. Thus, the relevant phases for
the space--time supersymmetry are the phases $c{S\choose{a_i}}$
where $a_i$ is any basis vector.
The sector $S$ produces
$N=4$ space--time supersymmetry and it is broken to $N=1$ by the
basis vectors $b_1$ and $b_2$. This breaking to $N=1$ is
irrespective of the choice of free phases $c{S\choose{b_1}}$ and
$c{S\choose{b_2}}$. However, once we fix those phases,
if we want to preserve $N=1$ space--time supersymmetry, all the remaining
$c{S\choose{a_i}}$ phases are fixed. For example, it is found that a
sufficient way to insure $N=1$ space--time supersymmetry is to impose
$c{S\choose{a_i}}=\delta_{a_i}$ (or $c{S\choose{a_i}}=-\delta_{a_i}$),
where $\delta_{a_i}=-1(+1)$ if $\psi^\mu$
periodic (antiperiodic) in the vector $a_i$, respectively. It is
noted that the first choice is the same as the phases
for the Neveu--Schwarz sector with any other vector,
$c{NS\choose{a_i}}=\delta_{a_i}$.

Therefore, with the first choice
to break supersymmetry at the Planck scale requires
that we take $c{S\choose{a_i}}\ne\delta_{a_i}$ for some basis
vector beyond the NAHE set. We could also choose
$c{S\choose{b_3}}\ne\delta_{b_3}$ but this is not a god choice
as it may also project out the fermions from the sector
$b_3$ rather than the bosons. Non--supersymmetric tachyon free
models can then be constructed. As an example we can
take the model of ref. \cite{eu}. Imposing
$c{S\choose{\alpha,\beta,\gamma}}\ne\delta_{\alpha,\beta,\gamma}$
produces a tachyon free non--supersymmetric model.
These models have in general a non--vanishing
cosmological constant and are therefore unstable due to the
presence of dilaton tadpole diagrams.

\subsection{Free phases beyond the NAHE set}

The GSO phases discussed in the previous sections
fix the spectrum that arises from the NAHE set basis vectors.
These are the free phases $c{b_j\choose b_i}$ $i,j=1,2,3$
and  $c{S\choose a_i}$, $a_i=\{S,b_1,b_2,b_3,\alpha,\beta,\gamma\}$.
In addition we have the freedom to choose, up to the modular invariance
constraints, the discrete phases
\beqn
&& c{b_j\choose{\alpha,\beta,\gamma}}~~~{\rm and}~~~\label{bjpbeyondnahe}\\
&& c{{\alpha,\beta,\gamma}\choose{\alpha,\beta,\gamma}}
\label{freephbeyondnahe}
\eeqn
These discrete choices of free phases affect as well
the physical spectrum and consequently
the low energy phenomenology for a given choice
of boundary condition basis vectors.
The phases in Eq. (\ref{bjpbeyondnahe}) affect mostly
the final charges of the states which are obtained from
the sectors $b_1$, $b_2$ and $b_3$ and therefore
fix the charges of three generations under the flavor
$U(1)$ symmetries. The phases in Eq. (\ref{freephbeyondnahe})
on the other hand fix the spectrum which arises
from the exotic sectors and give rise to the exotic
Wilsonian matter states. Therefore, these phases are
those which determine the spectrum in the models
which goes beyond the spectrum of the Standard Model.
As a first example the free phases will be shown to affect
the number of color triplets in the massless string spectrum
and therefore to play an important role in resolving
the string coupling unification problem.

It is well known that perturbative string unification predicts
that the gauge couplings unify at a scale which is about a
factor of twenty larger than the scale at which the couplings
are seen to intersect if one assumes only the MSSM spectrum below
the unification scale. This discrepancy seemingly should have
many possible resolutions remembering that it involves the extrapolation
of the gauge parameters over fifteen orders of magnitude.
However, surprisingly the problem is not easily resolved.
In ref. \cite{dienesfaraggi} it was shown that heavy string
threshold corrections, light SUSY thresholds, intermediate
enhanced non-Abelian gauge symmetry or modified weak
hypercharge normalization do not resolve the problem.
It was suggested that the only way to resolve the
problem is by having additional color triplets and electroweak
doublets beyond the spectrum of the minimal supersymmetric
standard model. Alternative proposals based on non--perturbative
effects \cite{witten,hsone} and product of moduli \cite{hstwo} were proposed.
However, it should be remarked that these proposals were made in
abstract setting and their realization in the context of
concrete, viable, string models remains an open question.

In some superstring models the extra needed representations
appear from sectors which arise due to the Wilson line breaking
of the non-Abelian gauge symmetries. The free phases play again
a crucial role in the determination of the physical spectrum
and therefore perform a very important function in fixing
the physical properties of the string models.

This essential role of the free phases is exemplified
with regard to the massless spectrum from the Wilsonian
sectors. As indicated above the Wilsonian sectors
produce additional massless states. These states can
be color triplets, electroweak doublets, color-weak
singlets with fractional electric charge, or they
can be Standard Model singlets with fractional
charge under the $U(1)_{Z^\prime}$. For example
in the model of ref. \cite{top}, we obtain several
additional electroweak doublets from the sectors
$b_1+b_3+\alpha\pm\gamma+(I)$ and $b_2+b_3+\beta\pm\gamma+(I)$,
while in the model of ref. \cite{gcu}
we obtain from the same sectors the color triplets
while the electroweak doublets are projected out.
The two models differ by the choice of one
and only one phase, with
\begin{equation}
c{\gamma\choose1}=-1 ~~~({\rm ref}. \cite{top})~~~\rightarrow~~~
c{\gamma\choose1}=+1 ~~~({\rm ref}. \cite{gcu})~~~
\label{topgcu}
\end{equation}

As a second example consider the model of ref. \cite{eu}.
In this model in addition to the Neveu--Schwarz sector,
the sector $X=b_1+b_2+b_3+\alpha+\beta+\gamma+(I)$, where
$I=1+b_1+b_2+b_3$, have $X_L\cdot X_L=0$ and $X_R\cdot X_R=8$.
Therefore, this sector may give rise to additional space--time
vectors bosons that would modify the space--time gauge group.
With the choice of GSO phases in ref. \cite{eu} all the extra gauge
bosons are projected out by the GSO projections. However,
with the modified GSO phases
\beq
c\left(\matrix{1\cr\gamma\cr}\right)\rightarrow
-c\left(\matrix{1\cr\gamma\cr}\right),
 c\left(\matrix{\alpha\cr\beta\cr}\right)\rightarrow
-c\left(\matrix{\alpha\cr\beta\cr}\right) ~\hbox{and}~
 c\left(\matrix{\gamma\cr\beta\cr}\right)\rightarrow
-c\left(\matrix{\gamma\cr\beta\cr}\right),
\label{euvectorbosons}
\eeq
additional space--time
vector bosons are obtained from the sector
$b_1+b_2+b_3+\alpha+\beta+\gamma+(I)$.
The sector $b_1+b_2+b_3+\alpha+\beta+\gamma+(I)$
produces the representations $3_1+3_{-1}$ of the hidden
$SU(3)_H$ gauge group, where
one of the $U(1)$ combinations is the $U(1)$ in the decomposition
of $SU(4)$ under $SU(3)\times U(1)$.
In this case the hidden $SU(3)_H$ gauge group is extended to
$SU(4)_H$ and the hidden sector contains two nonabelian factors
$SU(5)\times SU(4)$. The possibility of extending the hidden sector
gauge group from twisted sectors may be instrumental in trying to
implement the dilaton stabilization mechanism of ref. \cite{dv}.

As a further example of the role of the free phases, beyond the NAHE set,
in the determination of the phenomenology of the free fermionic models
consider the model in table [\ref{fnymodel}]
\beqn
 &\begin{tabular}{c|c|ccc|c|ccc|c}
 ~ & $\psi^\mu$ & $\chi^{12}$ & $\chi^{34}$ & $\chi^{56}$ &
        $\bar{\psi}^{1,...,5} $ &
        $\bar{\eta}^1 $&
        $\bar{\eta}^2 $&
        $\bar{\eta}^3 $&
        $\bar{\phi}^{1,...,8} $ \\
\hline
\hline
  ${b_4}$     &  1 & 1&0&0 & 1~1~1~1~1 & 1 & 0 & 0 & 0~0~0~0~0~0~0~0 \\
  ${\beta}$   &  1 & 0&0&1 & 1~1~1~0~0 & 1 & 0 & 1 & 1~1~1~1~0~0~0~0 \\
  ${\gamma}$  &  1 & 0&1&0 &
		${1\over2}$~${1\over2}$~${1\over2}$~${1\over2}$~${1\over2}$
	      & ${1\over2}$ & ${1\over2}$ & ${1\over2}$ &
                ${1\over2}$~0~1~1~${1\over2}$~${1\over2}$~${1\over2}$~1 \\
\end{tabular}
   \nonumber\\
   ~  &  ~ \nonumber\\
   ~  &  ~ \nonumber\\
     &\begin{tabular}{c|c|c|c}
 ~&   $y^3{y}^6$
      $y^4{\bar y}^4$
      $y^5{\bar y}^5$
      ${\bar y}^3{\bar y}^6$
  &   $y^1{\omega}^6$
      $y^2{\bar y}^2$
      $\omega^5{\bar\omega}^5$
      ${\bar y}^1{\bar\omega}^6$
  &   $\omega^1{\omega}^3$
      $\omega^2{\bar\omega}^2$
      $\omega^4{\bar\omega}^4$
      ${\bar\omega}^1{\bar\omega}^3$ \\
\hline
\hline
$\alpha$ & 1 ~~~ 0 ~~~ 0 ~~~ 1  & 0 ~~~ 0 ~~~ 1 ~~~ 0  & 0 ~~~ 0 ~~~ 1 ~~~ 0 \\
$\beta$  & 0 ~~~ 0 ~~~ 0 ~~~ 1  & 0 ~~~ 1 ~~~ 0 ~~~ 1  & 1 ~~~ 0 ~~~ 1 ~~~ 0 \\
$\gamma$ & 0 ~~~ 0 ~~~ 1 ~~~ 1  & 1 ~~~ 0 ~~~ 0 ~~~ 1  & 0 ~~~ 1 ~~~ 0 ~~~ 0 \\
\end{tabular}
\label{fnymodel}
\eeqn
With the choice of generalized GSO coefficients:
\beqn
c\left(\matrix{b_4\cr
                                    b_j,\beta\cr}\right)&&=
-c\left(\matrix{b_4\cr
                                    {\bf1}\cr}\right)=
-c\left(\matrix{\beta\cr
                                    {\bf1}\cr}\right)=
c\left(\matrix{\beta\cr
                                    b_j\cr}\right)=\nonumber\\
-c\left(\matrix{\beta\cr
                                    \gamma\cr}\right)&&=
c\left(\matrix{\gamma\cr
                                    b_2\cr}\right)=
-c\left(\matrix{\gamma\cr
                                    b_1,b_3,\alpha,\gamma\cr}\right)=
-1\nonumber
\eeqn
$(j=1,2,3),$
with the others specified by modular invariance and space--time
supersymmetry. With this choice of GSO projection coefficients
the following $U(1)$ symmetries are anomalous:
Tr${U_1=-24}$, Tr${U_2=-30}$, Tr${U_3=18}$,
Tr${U_5=6}$, Tr${U_6=6}$ and  Tr${U_8=12}$.
With the anomalous $U(1)$ symmetry being
\beq
U_A=-4U_1-5U_2+3U_3+U_5+U_6+2U_8.
\label{anomau1infny}
\eeq
Changing
\beq
c\left(\matrix{b_4\cr
                  1\cr}\right)=+1~\rightarrow~
            c\left(\matrix{b_4\cr
                  1\cr}\right)=-1,
\label{pb4tompb4}
\eeq
changes
the anomalous $U(1)$s to: Tr$U_C=-18$, Tr$U_L=12$,
Tr$U_1=-18$, Tr$U_2=-24$, Tr$U_3=24$,
Tr$U_4=-12$, Tr$U_5=6$, Tr$U_6=6$, Tr$U_7=-6$, Tr$U_8=12$
and Tr$U_9=18$, and the anomalous $U(1)$ combination is
\beq
U_A=-3U_C+2U_L-3U_1-4U_2+4U_3-2U_4+U_5+U_6-U_7+2U_8+3U_9.
\label{modfny}
\eeq
This modification has an important phenomenological implication.
In the model with the phase modification, Eq. (\ref{pb4tompb4}),
while the weak hypercharge combination,
$U_Y=1/3U_C+1/2U_L$, is anomaly free, the orthogonal
combination, which is embedded in $SO(10)$,
$U_{Z^\prime}=U_C-U_L$ is anomalous. This implies that there exist
models in which this $U_{Z^\prime}$ must be broken near the Planck
scale. In such models therefore the universal part of the
observable gauge group, arising from the $SO(10)$ gauge group of the
NAHE set, must be the Standard Model gauge group.
A $U(1)$ combination of the flavor dependent $U(1)s$, however,
may still remain unbroken.

The examples above illustrate the importance of the discrete choice
of the free phases in the phenomenology of the free fermionic models.
Depending on the choices of discrete phases we can construct models
with different phenomenological properties. Another
question of interest is how does the ``mirror symmetry'', exhibited
by the discrete change of phases in section (\ref{mirrorsec})
extend to the models that include basis vectors and
GSO phases beyond the NAHE set. The basis vectors beyond the
NAHE set correspond to Wilson lines in the orbifold language.
The duality symmetries in the extended fermionic models
then correspond to duality symmetries in orbifold constructions
that include Wilson lines. Naturally a complete classification
of all such duality symmetries is beyond the scope of this paper,
and in general it is difficult to assess what general, model independent,
patterns would persist. The simplest examples that can be considered
are the models that utilize solely periodic and antiperiodic
boundary conditions. These are the Pati--Salam $SO(6)\times SO(4)$
type models. As an example we can examine the model of the first
reference in \cite{alr}. The basis vectors and GSO phases are
shown in table [\ref{alrmodel}]
\beqn
 &\begin{tabular}{c|c|ccc|c|ccc|c}
 ~ & $\psi^\mu$ & $\chi^{12}$ & $\chi^{34}$ & $\chi^{56}$ &
        $\bar{\psi}^{1,...,5} $ &
        $\bar{\eta}^1 $&
        $\bar{\eta}^2 $&
        $\bar{\eta}^3 $&
        $\bar{\phi}^{1,...,8} $ \\
\hline
\hline
  ${b_4}$     &  1 & 1&0&0 & 1~1~1~1~1 & 1 & 0 & 0 & 0~0~0~0~0~0~0~0 \\
  ${b_5}$     &  1 & 0&1&0 & 1~1~1~1~1 & 0 & 1 & 0 & 0~0~0~0~0~0~0~0 \\
  $\alpha$     &  0 & 0&0&0 &
		1~1~1~0~0 & 1 & 1 & 1 & 0~0~1~1~0~0~0~0 \\
  ${\gamma}$ &  0 & 0&0&0 &
		1~1~1~1~1 & 1 & 1 & 1 & 1~1~1~1~0~0~0~0 \\
\end{tabular}
   \nonumber\\
   ~  &  ~ \nonumber\\
   ~  &  ~ \nonumber\\
     &\begin{tabular}{c|c|c|c}
 ~&   $y^3{\bar y}^3$
      $y^4{\bar y}^4$
      $y^5{\bar y}^5$
      ${y}^6{\bar y}^6$
  &   $y^1{\bar y}^6$
      $y^2{\bar y}^2$
      $\omega^5{\bar\omega}^5$
      ${\omega}^6{\bar\omega}^6$
  &   $\omega^2{\omega}^3$
      $\omega^1{\bar\omega}^1$
      $\omega^4{\bar\omega}^4$
      ${\bar\omega}^2{\bar\omega}^3$ \\
\hline
\hline
$b_4$    & 1 ~~~ 0 ~~~ 0 ~~~ 1  & 0 ~~~ 0 ~~~ 1 ~~~ 0  & 0 ~~~ 0 ~~~ 0 ~~~ 1 \\
$b_5$    & 0 ~~~ 0 ~~~ 0 ~~~ 1  & 0 ~~~ 1 ~~~ 1 ~~~ 0  & 0 ~~~ 1 ~~~ 0 ~~~ 0 \\
$\alpha$ & 0 ~~~ 1 ~~~ 0 ~~~ 1  & 0 ~~~ 0 ~~~ 0 ~~~ 1  & 0 ~~~ 0 ~~~ 1 ~~~ 1 \\
$\gamma$ & 0 ~~~ 0 ~~~ 0 ~~~ 1  & 0 ~~~ 0 ~~~ 0 ~~~ 1  & 0 ~~~ 0 ~~~ 0 ~~~ 0 \\
\end{tabular}
\label{alrmodel}
\eeqn
With the choice of generalized GSO coefficients:
\beqn
&& ~~c\left(\matrix{S\cr
                                    S,b_j,b_4,b_5\cr}\right)~=~
c\left(\matrix{b_2,b_3\cr
                                    b_4\cr}\right)~=~
c\left(\matrix{b_1,b_3\cr
                                    b_5\cr}\right)~=~
c\left(\matrix{b_4,b_5,\gamma\cr
                                    \gamma\cr}\right)~=~~~~\nonumber\\
&& ~~c\left(\matrix{b_1,b_2\cr
                                    \alpha\cr}\right)~=~
-c\left(\matrix{b_i\cr
                                    b_j\cr}\right)~=~
-c\left(\matrix{b_1,b_4\cr
                                    b_4\cr}\right)~=~
-c\left(\matrix{b_2,b_4,b_5\cr
                                    b_5\cr}\right)~=~~~~\nonumber\\
&& -c\left(\matrix{S,b_j\cr
                                    \gamma\cr}\right)~=~
-c\left(\matrix{S,b_1,b_4,b_5,b_6,\alpha\cr
                                    \alpha\cr}\right)~=~1
\label{alrphases}
\eeqn
$(j=1,2,3),$
with the others specified by modular invariance and space--time
supersymmetry. We can now examine what is the effect of discrete
modification of the GSO phases. Some modifications, as shown in the
examples above, will result in new models in which the physical
spectrum and the phenomenology is modified. However, certain
discrete changes will result in models in which the
spectrum is invariant but some of the charges under the $U(1)$
world--sheet currents will be modified. This is similar to
the mirror symmetry phenomena discussed in section (\ref{mirrorsec}).
Consider for example a global change of the GSO phases in the
[\ref{alrmodel}] model. Under this modification, as expected,
the chiral representations under $SO(10)$ from the sectors
$b_j$ $(j=1,2,3)$ reverse their chirality. Because of the
modification of the $c{S\choose{a_i}}$ GSO phases
the space--time gravitino reverses its chirality as well.
However the spectrum remains essentially identical.
Similarly, we can consider the model which preserves
the chirality of the space--time gravitino by keeping the
phases $c{S\choose{a_i}}$ unchanged, but taking the opposite
sign for all other phases. In this model again the chirality
of the chiral representations of the $SO(10)$ gauge group
is flipped, but up to this flip the models are identical.
Thus, there is a larger set of duality symmetries which
exist when considering the larger set of free phases
beyond the NAHE set and that may not be necessarily be
related to the geometrical dualities which have been discussed
in the literature.

\section{The $\alpha\beta$ sector}\label{ab}

In all the three generation free fermionic models, that utilize the NAHE set
and that were studied in detail to date,
there exist a special combination
of the basis vectors which extend the NAHE set. In the flipped
$SU(5)$ and in the Pati--Salam like models this combination
is the combination $S+b_4+b_5$, while in the standard--like models
of ref. \cite{eu} and ref. \cite{top} the combination is
$S+b_1+b_2+\alpha+\beta$. The $\alpha$, $\beta$ basis vectors
in these two models can be replaced by the combinations
$b_1+\alpha$ and $b_2+\beta$ which results, with a suitable
modification of the GSO phases, in the same physical spectrum.
The boundary conditions in this vector combination are, for example,
in the model of ref. \cite{eu},
$$\{\psi^{1,2},\chi^{56},\omega^1{\bar\omega}^1,y^2{\bar y}^2,
\omega^3{\bar\omega}^3,y^4{\bar y}^4,{\bar\eta}^1,{\bar\eta}^2\}
$$
are periodic while all the remaining world--sheet fermions are
antiperiodic. In this sector we have
$(\alpha\beta)_R\cdot(\alpha\beta)_R=4$ and
$(\alpha\beta)_L\cdot(\alpha\beta)_L=4$.
Massless states from this sector are obtained
by acting on the vacuum with one fermionic oscillator
with Neveu--Schwarz boundary conditions and
frequency $\pm1/2$. This sector then produces additional
vectorial representations of the observable $SO(10)$ symmetry
which after applying the GSO projections gives rise to
additional electroweak Higgs doublets and color triplets.
This is the important property of this sector as the
additional Higgs doublets from this sector are crucial
for obtaining qualitatively realistic fermion mass matrices.
In addition this vector combination produces additional
$SO(10)$ singlet fields which are charged under the
flavor $U(1)$ symmetries and are used in the cancelation of
the anomalous $U(1)$ D--term equation.
Therefore, it is desirable to require that the assignment
of boundary conditions beyond the NAHE set admits such
a vector combination in a given model. This requirement
in turn constrains further the possible distinct models.

\section{Classification by real fermion pairings}
One of the important constraints in the construction
of the free fermionic models is the requirement
of a well defined super--current. In the models that utilize
only periodic and anti--periodic boundary conditions
for the left--moving sector, the eighteen left--moving fermions
are divided into six triplets in the adjoint representation
of the automorphism group $SU(2)^6$.
These triplets are typically denoted by $\{\chi_i,y_i,\omega_i\}$
$i=1,\cdots,6$.
In this case
the allowed boundary conditions of each of these six triplets
depend on the boundary condition of the world--sheet fermions
$\psi_{1,2}^\mu$. For sectors with periodic boundary conditions,
$b(\psi^\mu_{1,2})=1$,
{\it i.e.} those that produce space time fermions the allowed
boundary condition in each triplet are $(1,0,0)$, $(0,1,0)$
$(0,0,1)$ and $(1,1,1)$. For sectors with antiperiodic
boundary conditions, $b(\psi^\mu_{1,2})=0$,
${\it i.e.}$ those that produce space--time
bosons, the allowed boundary conditions are $(1,1,0)$, $(1,0,1)$
$(0,1,1)$ and $(0,0,0)$. The left--moving fermions can be combined
in pairs to form
to form complex left--moving fermions. Alternatively,
a real left--moving fermion can be combined with a real
right--moving fermion to form an Ising model operator.
The super current constraint and the various desirable
phenomenological criteria then limit the possible
complex or Ising model combinations of the left--moving
fermions. Below I illustrate these constraints in several examples.

In the type of models that are considered here a pair of real
fermions which are combined to form a complex fermion
or an Ising model operator must have the identical boundary
conditions in all sectors. In practice it is sufficient
to require that a pair of such real fermions
have the same boundary conditions in all the boundary
basis vectors which span a given model. The NAHE set of
boundary condition basis vectors already divides the eighteen
left--moving real fermions into three groups

\beqn
&& \{(\chi^{1},~~~,~~~),(\chi^2,~~~,~~~),(~~~,y^{3},~~~),(~~~,y^4,~~~),
				  (~~~,y^5,~~~),(~~~,y^6,~~~)\}~~~~\\
&& \{(~~~,y^{1},~~~),(~~~,y^2,~~~),(\chi^{3},~~~,~~~),(\chi^{4},~~~,~~~),
			(~~~,~~~,\omega^5),(~~~,~~~,\omega^{6})\}~~~~\\
&& \{(~~~,~~~,\omega^1),(~~~,~~~,\omega^2),(~~~,~~~,\omega^3),
		(~~~,~~~,\omega^4),(\chi^{5},~~~,~~~),(\chi^6,~~~,~~~)\}~~~~
\label{reallmft}
\eeqn
where the notation emphasizes the division of the
eighteen left--moving internal world--sheet fermions
into the $SU(2)^6$ triplets.
The $\chi^{12,34,56}$ are the complexified combinations
which generate the $U(1)$ current of the $N=2$ left--moving
world--sheet supersymmetry \cite{kln}. We have the freedom
to complexify all, some or none of the remaining twelve
left--moving world--sheet fermions.
These different choices will in turn produce superstring
models with substantially different phenomenological
implications. The first obvious example is that if
all of the left--moving fermions are complexified
then the rank of the right--moving gauge group is
twenty--two, while if none are complexified then the
rank of right--moving gauge group is sixteen.

A more subtle example is noted from the discussion
in sections (\ref{higgs}) and (\ref{yukcop}).
The rules obtained there show that for the Higgs
color triplets to be projected out in the standard--like
models the condition $\Delta_j=1$ must hold, where
$\Delta_j$ is the difference between the boundary conditions
of the left-- and right--moving real world--sheet fermions.
Similarly the up--down quark Yukawa coupling
rule shows that to get a cubic level up type mass
term requires again $\Delta_j=1$. It follows that
if we impose the presence of Higgs doublets from the
Neveu--Schwarz sector in the massless string spectrum,
and consequently the projection of the corresponding color
Higgs triplets, that not all the twelve left--moving real
fermions can be combined to form Ising model operators,
as in this case we always have $\Delta_j=0$.
This example illustrate how the pairing
of the real world--sheet fermions affects the
phenomenology of the free fermionic models.

We can then ask further which
desired properties of the fermionic models, which are
motivated by the phenomenological constraints, can be
consistent with a given choice of pairings. Expressed
differently we can ask which properties are compatible
with a given choice of real fermions pairings and the
consistency of the world--sheet supercurrent.

As an example let us suppose that we would like to
impose the existence of three generations, the
existence of the type of sector discussed in section (\ref{ab}),
and the projection of all color Higgs triplets from the
Neveu--Schwarz sector and consider the pairing
\beqn
&& \{(y^3y^6,y^4{\bar y}^4,y^5{\bar y}^5,{\bar y^3}{\bar y}^6),\label{fnyp1}\\
&& (y^1\omega^6,y^2{\bar y}^2,\omega^5{\bar\omega}^5,
			{\bar y}^1{\bar\omega}^6),\label{fnyp2}\\
&& (\omega^1\omega^3,\omega^2{\bar\omega}^2,\omega^4{\bar\omega}^4,
				{\bar\omega^1}{\bar\omega}^3)\}
\label{fnypairing}
\eeqn
Note that with this pairing the complexified left-moving pairs
in Eqs. (\ref{fnyp1},\ref{fnyp2})
mix between the first, third and sixth $SU(2)$ triplets of the
left--moving automorphism group. The boundary condition of
a complexified pair then are forced to be identical in all boundary
condition basis vectors. The supercurrent constraints then restrict
the possible boundary conditions of the remaining real left--moving
world--sheet fermions. With this pairing we can start building the
phenomenological characteristics that we would like to impose.
We can start by constructing directly the $\alpha\beta$ sector
discussed in section (\ref{ab}), while at the same time
reducing the number of generations by a factor of $2$ from
each of the sectors $b_1$, $b_2$ and $b_3$. This is obtained
by the following assignment of boundary conditions
\beqn
     &\begin{tabular}{c|c|c|c}
 ~&   $y^3{y}^6$
      $y^4{\bar y}^4$
      $y^5{\bar y}^5$
      ${\bar y}^3{\bar y}^6$
  &   $y^1{\bar\omega}^6$
      $y^2{\bar y}^2$
      $\omega^5{\bar\omega}^5$
      ${\bar y}^1{\bar\omega}^6$
  &   $\omega^1{\omega}^3$
      $\omega^1{\bar\omega}^2$
      $\omega^4{\bar\omega}^4$
      ${\bar\omega}^1{\bar\omega}^3$ \\
\hline
\hline
$\alpha\beta$
	& 0 ~~~ 1 ~~~ 0 ~~~ 0  & 0 ~~~ 0 ~~~ 1 ~~~ 0  & 1 ~~~ 0 ~~~ 0 ~~~ 1 \\
\end{tabular}
\label{alphabetasector}
\eeqn
with $b_{\alpha\beta}\{\psi^\mu_{12},\chi^{56},
{\bar\eta}^1,{\bar\eta}^2\}=1$ as well.
Note that the modular invariance constraints on the product
$b_{\alpha\beta}\cdot b_j=0~{\rm mod}~2$ forces this
form of boundary conditions for the $\alpha\beta$ sector.
The next phenomenological criteria that we would like to
impose is that all color triplets from the Neveu--Schwarz
sector be projected by the GSO projections.  This requirement
forces the breaking of $SO(10)$ to $SO(6)\times SO(4)$, thus imposing
the boundary conditions $b\{{\bar\psi}^{1,\cdots,5}\}=\{1,1,1,0,0\}$.
Let us denote the additional basis vector as $\delta$.
In addition the quantity $\Delta_j$ from section (\ref{higgs})
must be equal to one for $(j=1,2,3)$. Without loss of generality
it is sufficient to contemplate possible assignments with
$b_\delta\{\psi^\mu_{12},\chi^{12},\chi^{34},\chi^{56}\}=0$ as
all other allowed possibilities can be obtained with combinations
of the NAHE set basis vectors.
Note that the basis vector $\delta$ can only contain
periodic and anti--periodic boundary conditions.
If we assign $1/2$ boundary condition to
${\bar\phi}^{1,\cdots,8}$ we then have $2\delta={{\bf1}+b_1+b_2+b_3}$
in contradiction with the rule that all basis vectors must
be linearly independent. The modular invariance
restrictions on $\delta\cdot b_j=0~{\rm mod}~2$ then forces
the vector $\delta$ to have the form
\beqn
     &\begin{tabular}{c|c|c|c}
 ~&   $y^3{y}^6$
      $y^4{\bar y}^4$
      $y^5{\bar y}^5$
      ${\bar y}^3{\bar y}^6$
  &   $y^1{\bar\omega}^6$
      $y^2{\bar y}^2$
      $\omega^5{\bar\omega}^5$
      ${\bar y}^1{\bar\omega}^6$
  &   $\omega^1{\omega}^3$
      $\omega^1{\bar\omega}^2$
      $\omega^4{\bar\omega}^4$
      ${\bar\omega}^1{\bar\omega}^3$ \\
\hline
\hline
$\delta$
	& 1~~~ 1 ~~~ 0 ~~~ 0  & 1 ~~~ 1 ~~~ 0 ~~~ 0  & 1 ~~~ 1 ~~~ 1 ~~~ 0 \\
\end{tabular}
\label{deltavector}
\eeqn
However, this form does not satisfy the modular invariance
constraints on the product $b_\delta\cdot b_{\alpha\beta}$.
It is therefore found that this two phenomenological properties,
namely existence of $\alpha\beta$ sector and projection
of color triplets from the Neveu--Schwarz sector cannot be
mutually compatible with this choice of pairing of the left--moving
real world--sheet fermions.

Next let us consider the pairings
\beqn
&& \{(y^3y^6,y^4{\bar y}^4,y^5{\bar y}^5,{\bar y^3}{\bar y}^6), \nonumber\\
&& (y^1\omega^6,y^2{\bar y}^2,\omega^5{\bar\omega}^5,
			{\bar y}^1{\bar\omega}^6),\nonumber\\
&& (\omega^2\omega^4,\omega^1{\bar\omega}^1,\omega^3{\bar\omega}^3,
				{\bar\omega^2}{\bar\omega}^4)\}
\label{newpairing}
\eeqn
Note that with this pairing the complexified left-moving pairs
mix between the sixth $SU(2)$ triplet of the
left--moving automorphism group. That is the boundary condition
of $y^3y^6$ fixes the boundary condition of $y^1\omega^6$.
A basis vector of the form of $b_{\alpha\beta}$ can
again be constructed directly by the following assignment
of boundary conditions
\beqn
     &\begin{tabular}{c|c|c|c}
 ~&   $y^3{y}^6$
      $y^4{\bar y}^4$
      $y^5{\bar y}^5$
      ${\bar y}^3{\bar y}^6$
  &   $y^1{\bar\omega}^6$
      $y^2{\bar y}^2$
      $\omega^5{\bar\omega}^5$
      ${\bar y}^1{\bar\omega}^6$
  &   $\omega^2{\omega}^4$
      $\omega^1{\bar\omega}^2$
      $\omega^3{\bar\omega}^3$
      ${\bar\omega}^2{\bar\omega}^4$ \\
\hline
\hline
$\alpha\beta$
	& 0 ~~~ 1 ~~~ 0 ~~~ 0  & 0 ~~~ 0 ~~~ 1 ~~~ 0  & 0 ~~~ 1 ~~~ 1 ~~~ 0 \\
\end{tabular}
\label{newalphabeta}
\eeqn
with $b\{\psi^\mu_{12},\chi^{56},{\bar\eta}^1,{\bar\eta}^2\}=1$ as well.
The projection of the color triplet Higgs from the Neveu--Schwarz
sector is achieved with $b\{{\bar\psi}^{1,\cdots,5}\}=\{1,1,1,0,0\}$.
Again without loss of generality
it is sufficient to contemplate possible assignments with
$b\{\psi^\mu_{12},\chi^{12},\chi^{34},\chi^{56}\}=0$ as
all other allowed possibilities can be obtained with combinations
of the NAHE set basis vectors.
The modular invariance
restrictions on $\delta\cdot b_j=0~{\rm mod}~2$ then forces
the vector $\delta$ to have the form
\beqn
     &\begin{tabular}{c|c|c|c}
 ~&   $y^3{y}^6$
      $y^4{\bar y}^4$
      $y^5{\bar y}^5$
      ${\bar y}^3{\bar y}^6$
  &   $y^1{\bar\omega}^6$
      $y^2{\bar y}^2$
      $\omega^5{\bar\omega}^5$
      ${\bar y}^1{\bar\omega}^6$
  &   $\omega^2{\omega}^4$
      $\omega^1{\bar\omega}^2$
      $\omega^3{\bar\omega}^3$
      ${\bar\omega}^2{\bar\omega}^4$ \\
\hline
\hline
$\delta$
	& 0~~~ 1 ~~~ 0 ~~~ 1  & 0 ~~~ 1 ~~~ 0 ~~~ 1  & 1 ~~~ 0 ~~~ 0 ~~~ 0 \\
\end{tabular}
\label{newdeltavector}
\eeqn
and the remaining string consistency constraints are fixed
by choosing appropriate boundary conditions for the world--sheet
fermions which generate the hidden gauge group.
Therefore, we note that both the existence of a $\alpha\beta$
combination as well as the projection of the color Higgs triplets
can be compatible with this choice of pairing of the left--moving
world--sheet fermions. However, the number of generations must still
be reduced by breaking the degeneracy among the world--sheet fermions
$\{(y^3y^6,y^5{\bar y}^5),(y^1\omega^6,\omega^5{\bar\omega}^5),
(\omega^1{\bar\omega}^1,\omega^3{\bar\omega}^3)\}$.
We must therefore assign either $b_\gamma(\omega^1)=1$ or
$b_\gamma(\omega^3)=1$ but not both.
Again I am considering assignments with
$b_\gamma\{\psi^\mu_{12},\chi^{12},\chi^{34},\chi^{56}\}=0$.
We note however that
$b_\gamma(\omega^1)=1$ and the supercurrent constraint
forces $b_\gamma(y^1,\omega^6)=1$ which forces
$b_\gamma(y^3,y^6)=1$ which forces $b_\gamma(\omega^3)=1$.
Similarly, we can consider the
other possible assignments for the complex world--sheet
fermions $\{\psi^\mu_{12},\chi^{12},\chi^{34},\chi^{56}\}$
observing that the degeneracy cannot be removed in any case.
Therefore, the degeneracy between
$\{(\omega^1{\bar\omega}^1,\omega^3{\bar\omega}^3)\}$.
cannot be removed and a three generation model, compatible
with these phenomenological criteria and this choice of
pairings cannot be obtained.

Next let us consider the pairings
\beqn
&& \{(y^3y^6,y^4{\bar y}^4,y^5{\bar y}^5,{\bar y^3}{\bar y}^6), \nonumber\\
&& (y^1\omega^6,y^2{\bar y}^2,\omega^5{\bar\omega}^5,
			{\bar y}^1{\bar\omega}^6),\nonumber\\
&& (\omega^1\omega^4,\omega^2{\bar\omega}^2,\omega^3{\bar\omega}^3,
				{\bar\omega^1}{\bar\omega}^4)\},
\label{w1w4pairing}
\eeqn
and consider the formation of a $\alpha\beta$ sector.
A sector $\alpha\beta$ is symmetric with respect
to two of the NAHE set basis vectors $b_j$. Because of
the cyclic permutation of the NAHE set we can impose, without
loss of generality, that the $\alpha\beta$ sector be symmetric
with respect to $b_1$ and $b_2$. This means that in the vector
$\alpha\beta$ the world--sheet fermions $\{\psi^{\mu}_{12},\chi^{56}\vert
{\bar\eta}^1,{\bar\eta}^2\}$ have periodic boundary conditions.
We observe that with the choice made in Eq. (\ref{w1w4pairing})
a $\alpha\beta$ vector which is symmetric with respect
to $b_1$ and $b_2$ cannot be formed. This follows because
each one of the first four $SU(2)$ triplets of the left--moving
supercurrent must contain one and only one periodic fermion.
On the other hand, to reduce the number of generations,
we impose that each of the sets in Eq. (\ref{w1w4pairing})
have at least one periodic fermion in the $\alpha\beta$ vector.
This two requirements are observed not to be mutually compatible.
A similar consideration also applies to the choice of
pairings
\beqn
&& \{(y^3y^6,y^4{\bar y}^4,y^5{\bar y}^5,{\bar y^3}{\bar y}^6), \nonumber\\
&& (y^1\omega^5,y^2{\bar y}^2,\omega^6{\bar\omega}^6,
			{\bar y}^1{\bar\omega}^5),\nonumber\\
&& (\omega^1\omega^4,\omega^2{\bar\omega}^2,\omega^3{\bar\omega}^3,
				{\bar\omega^1}{\bar\omega}^4)\}.
\label{y3y6y1w5w1w4}
\eeqn

Next let us consider the pairing
\beqn
&& \{(y^3y^6,y^4{\bar y}^4,y^5{\bar y}^5,{\bar y^3}{\bar y}^6), \nonumber\\
&& (y^1\omega^5,y^2{\bar y}^2,\omega^6{\bar\omega}^6,
			{\bar y}^1{\bar\omega}^5),\nonumber\\
&& (\omega^2\omega^4,\omega^1{\bar\omega}^1,\omega^3{\bar\omega}^3,
				{\bar\omega^2}{\bar\omega}^4)\}.
\label{278pairing}
\eeqn
This pairing mixes all six triplets of the $SU(2)^6$
left--moving automorphism group and is the one used in
the models of refs. \cite{eu} and \cite{top}.
With this pairing the following assignment of boundary
conditions satisfies the desired phenomenological criteria,
namely, three generations, existence of $\alpha\beta$ sector
and projection of the color Higgs triplets from the Neveu--Schwarz
sector by, for example, the following assignment of boundary
conditions,
\beqn
     &\begin{tabular}{c|c|c|c}
 ~&   $y^3{y}^6$
      $y^4{\bar y}^4$
      $y^5{\bar y}^5$
      ${\bar y}^3{\bar y}^6$
  &   $y^1{\bar\omega}^5$
      $y^2{\bar y}^2$
      $\omega^6{\bar\omega}^6$
      ${\bar y}^1{\bar\omega}^5$
  &   $\omega^2{\omega}^4$
      $\omega^1{\bar\omega}^2$
      $\omega^3{\bar\omega}^3$
      ${\bar\omega}^2{\bar\omega}^4$ \\
\hline
\hline
$\alpha\beta$
	& 0~~~ 1 ~~~ 0 ~~~ 0  & 0 ~~~ 1 ~~~ 0 ~~~ 0  & 0 ~~~ 1 ~~~ 1 ~~~ 0 \\
$\delta$
	& 1~~~ 0 ~~~ 0 ~~~ 0  & 0 ~~~ 0 ~~~ 1 ~~~ 1  & 0 ~~~ 0 ~~~ 1 ~~~ 1 \\
$\gamma$
	& 0~~~ 1 ~~~ 0 ~~~ 1  & 0 ~~~ 1 ~~~ 0 ~~~ 1  & 1 ~~~ 0 ~~~ 0 ~~~ 0 \\
\end{tabular}
\label{m278274}
\eeqn
The above examples illustrate the effect the different
choices of pairings of the left--moving
real world--sheet fermions
on the massless spectrum and  phenomenological properties
of the free fermionic models. Clearly the choices considered
here do not exhaust all the possibilities and it is of further
interest to examine which pairings can be compatible with
the array of phenomenological constraints that must be
imposed on a realistic superstring model.

An interesting question is what is the number of
independent pairings that can be formed. Before imposing
the NAHE set boundary conditions we have twelve left--moving
real fermions. Any of those can be combined to form Ising
model or complex fermion. After imposing the NAHE set projections
the twelve left--moving real fermions are divided into
three groups of four left--moving real fermions each.
Thus, the number of allowed pairings is reduce significantly.
{}From each group of four real fermions
we can form ten possible combinations.
For example for the set $\{y^{3,\cdots,6}\}$ we can form
the combinations
\beqn
&& (y^3y^4~,~y^5y^6)~~,~~
   (y^3y^4~,~y^5{\bar y}^5~,~y^6{\bar y}^6)~~,~~
   (y^3{\bar y}^3~,~y^4{\bar y}^4~,~y^5y^6)~~,\\
&& (y^3y^5~,~y^4y^6)~~,~~
   (y^3y^5~,~y^4{\bar y}^4~,~y^6{\bar y}^6)~~,~~
   (y^3{\bar y}^3~,~y^5{\bar y}^5~,~y^4y^6)~~,\\
&& (y^3y^6~,~y^4y^5)~~,~~
   (y^3y^6~,~y^4{\bar y}^4~,~y^5{\bar y}^5)~~,~~
   (y^3{\bar y}^3~,~y^6{\bar y}^6~,~y^4y^5)~~,~~\\
&& (y^3{\bar y}^3~,~y^4{\bar y}^4~,~y^5{\bar y}^5~,~y^6{\bar y}^6)~~.
\label{pairingsy36}
\eeqn
With similar number of possible
pairings from the sets $\{y^{1,2},\omega^{5,6}\}$
and $\{\omega^{1,\cdots,4}\}$. Thus, we have a total of one thousand
possible pairings. Due to the permutation symmetry of
the NAHE set and a $Z_2$ symmetry due to the exchange
of the two triplets in each set, we can reduce this number
by at least a factor of six. We are still left with a
number of pairings in excess of 150 possibilities,
some of which may still equivalent. However, this illustrates
the large number of possible distinct models. It is expected
that all possible pairings can produce three generation models
with either one of the $SO(10)$ subgroups and the distinct
choices differ by their other phenomenological properties
as exemplified above.

\section{Classification by exotics}

In the free fermionic models which utilize the NAHE
set the structure of the Standard Model fermion sector
is obtained from the NAHE set basis vectors and possibly
from one or two of the basis vectors which extend the NAHE
set. Therefore the structure of the observable matter
spectrum is similar, up to $U(1)$ charges, in different
models. The models differ only by the basis vectors
which extend the NAHE set. These basis vectors and their
combinations with the NAHE set basis vectors produce
additional massless spectrum. In the flipped $SU(5)$ and
$SO(6)\times SO(4)$ models two of the basis vectors,
denoted $b_4$ and $b_5$, do not break the $SO(10)$
symmetry and produce additional $SO(10)$ spinorial
representations. However, in each class of models
at least one of the basis vectors must break the
$SO(10)$ symmetry. Such a basis vector, combined
with its combinations with the NAHE set basis vectors,
produces additional matter states which do not fall
into representations of the original $SO(10)$ gauge
group. These new states arise due to the breaking
of the GUT gauge group at the string level rather
than at the effective field theory level. They are
a characteristic of the breaking of the non-Abelian
gauge symmetry by Wilson lines and are a generic
feature of superstring compactifications
\cite{fcothers,ccf}. In the realistic free fermionic
models these type of exotic states can be classified
first according to the sectors in which they appear.
Namely in each of the sectors which produces exotic
states we have different boundary conditions for the
right--moving world--sheet fermions,
${\bar\psi}^{1,\cdots,5}$, which generate the $SO(10)$
symmetry. Each type of boundary conditions
results in different types of exotic states.
Each choice of the final $SO(10)$ gauge subgroup,
$SU(5)\times U(1)$, $SO(6)\times SO(4)$ or
$SU3)\times SU(2)\times U(1)^2$ allows distinct
types of exotic states. The last possibility,
as it contains the previous two $SO(10)$
breaking sectors, admits exotic states
which appear also in the $SU(5)\times U(1)$
and $SO(6)\times SO(4)$ models, and in addition
gives rise to exotic states which can appear
only in the $SU(3)\times SU(2)\times U(1)^2$
models. Thus, one classification by the exotic states
is by the type of final $SO(10)$ subgroup which
is left unbroken. The other possible classification
classifies the models by the actual exotic
representations which appear in each model.

The $SO(6)\times SO(4)$ type exotic states
are obtained from sectors with $X_R\cdot X_R=8$
and with
\beq
X({\bar\psi}^{1,\cdots,5})=(1,1,1,0,0)
\label{so64et}
\eeq
or
\beq
X({\bar\psi}^{1,\cdots,5})=(0,0,0,1,1).
\label{so64ed}
\eeq
The $SU(5)\times U(1)$ type exotic states
may be produced from sectors with $X_R\cdot X_R=8$,
$X_R\cdot X_R=6$ or $X_R\cdot X_R=4$
and with
\beq
X({\bar\psi}^{1,\cdots,5})=({1\over2},{1\over2},{1\over2},{1\over2},{1\over2})
\label{su5u1e}
\eeq
The $SU(3)\times SU(2)\times U(1)^2$ type exotic states
may be produced from sectors with $X_R\cdot X_R=8$,
$X_R\cdot X_R=6$ or $X_R\cdot X_R=4$ and with
\beq
X({\bar\psi}^{1,\cdots,5})=({1\over2},{1\over2},{1\over2},
					-{1\over2},-{1\over2})
\label{su321e}
\eeq
where it should remembered that the weak hypercharge
is defined by $$U(1)_Y={1\over3}U(1)_C+{1\over2}U(1)_L.$$
In the following I use the notation
\begin{equation}
[(SU(3)_C\times U(1)_C);
     (SU(2)_L\times U(1)_L)]_{(Q_Y,Q_{Z^\prime},Q_{\rm e.m.})},
\label{notation}
\end{equation}
where $Q_Y$ is the weak hypercharge of a given state,
$Q_{Z^\prime}$ is the charge under the orthogonal combination
$U(1)_{Z^\prime}=U(1)_C-U(1)_L$, and for the electroweak doublets the
electric charge of the two components is written.

The possible $SO(6)\times SO(4)$ type exotic states are
\beqn
&&[(    3, {1\over2});(1,0)]_{( 1/6, 1/2, 1/6)}~~~~;~~~~
[({\bar3},-{1\over2});(1,0)]_{(-1/6,-1/2,-1/6)}~~~~;\\
&&[(1,\pm3/2);(1,0)]_{(\pm1/2,\pm1/2,\pm1/2)}
\label{so64etes}
\eeqn
from sectors of the form of Eq. (\ref{so64et}) and
\beqn
&&[(1,0);(2,0)]_{(0,0,\pm1/2)}\\
&&[(1,0);(1,\pm{1})]_{(\pm1/2,\mp1/2,\pm1/2)}
\label{so64edes}
\eeqn
from sectors of the form of Eq. (\ref{so64ed}).

The possible $SU(5)\times U(1)$ type exotic states are
\beq
[(1,\pm3/4);(1,\pm{1/2})]_{(\pm1/2,\pm1/4,\pm1/2)}
\label{fc51singlet}
\eeq
from sectors of the form of Eq. (\ref{su5u1e}) with $X_R\cdot X_R=8$.
In the other two cases, $X_R\cdot X_R=6$ and $X_R\cdot X_R=4$,
the massless states are obtained acting on the
vacuum with one, or two, fermionic oscillators with frequency
$1/4$, respectively. We can then obtain states that transform
as $(5,1/4)$, $({\bar5},-1/4)$, $(10,1/2)$, $({\overline{10}},-1/2)$,
under $SU(5)\times U(1)$.

In $SU(3)\times SU(2)\times U(1)^2$ type models there may exist
sectors of the $SO(6)\times SO(4)$ and $SU(5)\times U(1)$
types, and in addition sectors of the form of Eq. (\ref{su321e}) which
can appear only in the $SU(3)\times SU(2)\times U(1)^2$ models.
These sectors are obtained from combinations of the
basis vectors that contain both the $SO(6)\times SO(4)$ and
$SU(5)\times U(1)$ breaking vectors.
The possible $SU(3)\times SU(2)\times U(1)^2$ type exotic states are
\beqn
&&[(3,{1\over4});(1,{1\over2})]_{(-1/3,-1/4,-1/3)}~~~~;~~~~
[(\bar3,-{1\over4});(1,{1\over2})]_{(1/3,1/4,1/3)}~~~~;\label{uniton}\\
&&[(1,\pm{3\over4});(2,\pm{1\over2})]_
			{(\pm1/2,\pm1/4,(1,0);(0,-1))}~~~~;\label{s321ed}\\
&&[(1,\pm{3\over4});(1,\mp{1\over2})]_{(0,\pm5/4,0)},\label{s321es}
\eeqn
Sectors with $X_R\cdot X_R=8$ give rise only to exotic states
which are Standard Model singlets with the quantum numbers
of Eq. (\ref{s321es}). Sectors with $X_R\cdot X_R=6$ give
rise to all the exotic states with the quantum numbers of
Eqs. (\ref{uniton},\ref{s321ed},\ref{s321es}). In addition to the
states above, sectors with $X_R\cdot X_R=4$ can give rise
to exotic diquarks. In these sectors the massless states are
obtained by acting on the vacuum with two fermionic oscillators
with frequency $1/4$, thus producing exotic diquarks with
\begin{equation}
[(3,-{1\over4}),(2,{1\over2})]_{(1/6,-3/4,(2/3,-1/3)}~~;~~
[(\bar3,{1\over4}),(2,{1\over2})]_{(-1/6,3/4,(-2/3,1/3)} \ \ .
\label{exoticdq2}
\end{equation}
from sectors with the boundary conditions in Eq. (\ref{su321e})
while sectors with the boundary conditions in Eq. (\ref{su5u1e})
with $X_R\cdot X_R=4$ may give rise to exotic diquarks with
\begin{equation}
[(3,-{1\over4}),(2,-{1\over2})]_{(-1/3,1/4,(1/6,-5/6)}~~;~~
[(\bar3,{1\over4}),(2,{1\over2})]_{(1/3,-1/4,(-1/6,5/6)} \ \ ,
\label{exoticdq1}
\end{equation}

The above classification exhausts all the possible exotic
states which may in principle appear in the realistic free
fermionic models. The classification then proceeds by examining
which type of sectors and therefore which type of exotic states
actually appear in specific models. The task then is to further
understand which choices of boundary condition basis vectors
and generalized GSO projection coefficients produce
specific classes of exotic states.

Such a complete classification and understanding is beyond
the scope of the present paper. However, short of that, we
can still make several important observations.
The first observations is in regard to the presence of
states with fractional electric charge.
Both the $SU(5)\times U(1)$ and $SO(6)\times SO(4)$ type
exotic states carry electric charge $\pm1/2$.
{}From a phenomenological point of view such states must
be confined to produce integrally charged states or
must become super--heavy by the breaking of the four
dimensional gauge group along the flat F and D directions.
In the revamped flipped $SU(5)$ model \cite{revamp} it is
found that all the fractionally charged states transform
in representations of the hidden non--Abelian gauge groups.
Therefore, in this model indeed all the fractionally charged
states are confined. As an example of the alternative solution,
examining the fractionally charged states and the
cubic level superpotential of the model of
ref. \cite{fny}
\begin{eqnarray}
W_2&=&{1\over{\sqrt2}}\{H_1H_2\phi_4+H_3H_4{\bar\phi}_4+
H_5H_6{\bar\phi}_4+(H_7H_8+H_9H_{10})\phi_4'\nonumber\\
&+&(H_{11}+H_{12})(H_{13}+H_{14}){\bar\phi}_4'
 +V_{41}V_{42}{\bar\phi}_4+V_{43}V_{44}{\bar\phi}_4\nonumber\\
&+&V_{45}V_{46}\phi_4+(V_{47}V_{48}+V_{49}V_{50}){\bar\phi}_4'
+V_{51}V_{52}\phi_4'\}\nonumber\\
&+&[H_{15}H_{16}\phi_{56}'
+H_{17}H_{18}{\bar\phi}_{56}+H_{19}H_{20}{\bar\phi}_{56}'+
H_{21}H_{22}{\bar\phi}_{56}\nonumber\\
&+&(V_{11}V_{12}+V_{13}V_{14})\phi_{13}+
(V_{15}+V_{16})(V_{17}+V_{18})\phi_{13}+V_{19}V_{20}\phi_{13}\nonumber\\
&+&V_{21}V_{22}\phi_{12}+V_{23}V_{24}\phi_{12}+
(V_{25}+V_{26})(V_{27}+V_{28})\phi_{12}+V_{29}V_{30}\phi_{12}\nonumber\\
&+&V_{31}V_{32}{\bar\phi}_{23}
+V_{33}V_{34}\phi_{23}+H_{29}H_{30}{\bar\phi}_{13}+
H_{36}H_{37}\phi_{12}],\label{fnysuperp}
\end{eqnarray}
it is observed that all the fractionally charged
states receive a Planck scale mass by giving a VEV to
the neutral singlets
${\bar\phi}_4,{\bar\phi}_4',{\phi}_4,\phi_4'$
which imposes  the additional F flatness constraint
$(\phi_4{\bar\phi}_4'+{\bar\phi}_4\phi_4')=0$. The other exotic states
which are Standard Model singlets do not receive mass by this choice
of flat direction. Therefore, at this level of the superpotential, the
fractionally charged states can decouple from the remaining light
spectrum. This example indicates the existence of free fermionic
models in which fractionally charged states appear only at the
massive level.

The question of further interest is therefore which exotic
sectors and corresponding states actually arise in given models.
One example is provided by the state in Eq. (\ref{uniton}).
This state is a color triplet and is obtained from
$SU(3)\times SU(2)\times U(1)^2$ type exotic sectors
with $X_R\cdot X_R=6$, which requires one fermionic
oscillator acting on the vacuum. The states from this
type of sectors are of particular phenomenological
interest because they carry the standard charges under
the Standard Model gauge group, but carry ``fractional''
charges under the $U(1)_{Z^\prime}$ embedded in $SO(10)$.
Consequently, this type of matter states may in fact
remain sufficiently light and be in agreement with current
experimental constraints. Such sectors and states appear
generically in the free fermionic Standard--like models.
The question then is whether their appearance is a necessary
property of this choice of vacuum. The following model illustrates
that this is not the case
\beqn
     &\begin{tabular}{c|c|c|c}
 ~&   $y^3{y}^6$
      $y^4{\bar y}^4$
      $y^5{\bar y}^5$
      ${\bar y}^3{\bar y}^6$
  &   $y^1{\bar\omega}^5$
      $y^2{\bar y}^2$
      $\omega^6{\bar\omega}^6$
      ${\bar y}^1{\bar\omega}^5$
  &   $\omega^2{\omega}^4$
      $\omega^1{\bar\omega}^2$
      $\omega^3{\bar\omega}^3$
      ${\bar\omega}^2{\bar\omega}^4$ \\
\hline
\hline
$\alpha$
	& 1~~~ 0 ~~~ 0 ~~~ 0  & 0 ~~~ 0 ~~~ 1 ~~~ 1  & 0 ~~~ 0 ~~~ 1 ~~~ 0 \\
$\beta$
	& 0~~~ 0 ~~~ 1 ~~~ 1  & 1 ~~~ 0 ~~~ 0 ~~~ 0  & 0 ~~~ 1 ~~~ 0 ~~~ 0 \\
$\gamma$
	& 0~~~ 1 ~~~ 0 ~~~ 1  & 0 ~~~ 1 ~~~ 0 ~~~ 1  & 1 ~~~ 0 ~~~ 0 ~~~ 0 \\
\end{tabular}
\label{m278modified}
\eeqn
With the choice of generalized GSO projections coefficients:
\beqn
&& c\left(\matrix{b_i\cr
                          S,b_j,\alpha,\beta,\gamma\cr}\right)=
c\left(\matrix{\alpha,\beta,\gamma\cr
                                    \alpha\cr}\right)=
c\left(\matrix{\beta\cr
                                    \beta\cr}\right)=\nonumber\\
&& c\left(\matrix{\gamma\cr
                                    {\bf1}\cr}\right)=
-c\left(\matrix{\beta\cr
                                    \gamma\cr}\right)=-1\nonumber
\eeqn
with $(j=1,2,3)$.
This model is a modification of the model of ref. \cite{eu}
obtained by taking $\alpha({\bar\omega}^2{\bar\omega}^4)=1\rightarrow
\alpha({\bar\omega}^2{\bar\omega}^4)=0$ and
$\beta({\bar\omega}^2{\bar\omega}^4)=1\rightarrow
\beta({\bar\omega}^2{\bar\omega}^4)=0$. With this modification and
the rules of section (\ref{higgs}) we see that in this model
we obtain from the Neveu--Schwarz sector the Higgs color triplets
$D_3$ and ${\bar D}_3$, rather than the corresponding doublets
$h_3$ and ${\bar h}_3$. Spanning the additive group, with this
choice of boundary condition basis vectors, it is observed that
this model does not contain any sector with $X_R\cdot X_R=6$.
Therefore, this model does not contain states from this
type of exotic sectors irrespective of the choice of
the generalized GSO projection coefficients.
This shows the existence of standard--like
models in which the states of the form of Eq. (\ref{uniton}) do not appear.
Another example is provided by the revamped flipped
$SU(5)$ model in which sectors with $X_R\cdot X_R=6$ do appear.
In the flipped $SU(5)$ this type of sectors generate exotic
$5$ and $\bar 5$ representations with $U(1)_5$ charges $\pm1/4$.
However, there the exotic $5$ and ${\bar5}$ are projected
out by the GSO projections. Therefore, for a given exotic state
to exist in a specific model, the type of exotic sector must
appear in the additive group and the state must not be projected
out by the GSO projections.

As another example of the classification by exotics
we note that none of the three generation free fermionic
models, which were studied in detail to date, contains
sectors that can produce diquark states.
However, such a model can be constructed by building the
desired vector with $X_R\cdot X_R=4$ into the set of basis
vectors. Table [\ref{diquarkmodel}] provides an example of such
a model.
\beqn
 &\begin{tabular}{c|c|ccc|c|ccc|c}
 ~ & $\psi^\mu$ & $\chi^{12}$ & $\chi^{34}$ & $\chi^{56}$ &
        $\bar{\psi}^{1,...,5} $ &
        $\bar{\eta}^1 $&
        $\bar{\eta}^2 $&
        $\bar{\eta}^3 $&
        $\bar{\phi}^{1,...,8} $ \\
\hline
\hline
  $\alpha$     &  1 & 1&0&0 & 1~1~1~0~0 & 0 & 0 & 1 & 1~1~0~0~0~0~0~0 \\
  $\beta$     &  1 & 0&1&0 & 1~1~1~0~0 & 0 & 0 & 1 & 1~1~0~0~0~0~0~0 \\
  ${\gamma}$  &  1 & 0&0&1 &
		${1\over2}$~${1\over2}$~${1\over2}$~${1\over2}$~${1\over2}$
		& ${1\over2}$ & ${1\over2}$ & ${1\over2}$ &
                0~0~0~0~${1\over2}$~${1\over2}$~${1\over2}$~${1\over2}$ \\
\end{tabular}
   \nonumber\\
   ~  &  ~ \nonumber\\
   ~  &  ~ \nonumber\\
     &\begin{tabular}{c|c|c|c}
 ~&   $y^3{\bar y}^3$
      $y^4{\bar y}^4$
      $y^5{\bar y}^5$
      ${y}^6{\bar y}^6$
  &   $y^1{\bar\omega}^6$
      $y^2{\bar y}^2$
      $\omega^5{\bar\omega}^5$
      ${\bar y}^1{\bar\omega}^6$
  &   $\omega^1{\omega}^3$
      $\omega^2{\bar\omega}^2$
      $\omega^4{\bar\omega}^4$
      ${\bar\omega}^1{\bar\omega}^3$ \\
\hline
\hline
$\alpha$ & 1 ~~~ 0 ~~~ 0 ~~~ 0  & 0 ~~~ 0 ~~~ 1 ~~~ 0  & 0 ~~~ 0 ~~~ 1 ~~~ 1 \\
$\beta$  & 0 ~~~ 0 ~~~ 1 ~~~ 0  & 1 ~~~ 0 ~~~ 0 ~~~ 0  & 0 ~~~ 1 ~~~ 0 ~~~ 0 \\
$\gamma$ & 0 ~~~ 1 ~~~ 0 ~~~ 0  & 0 ~~~ 1 ~~~ 0 ~~~ 0  & 1 ~~~ 0 ~~~ 0 ~~~ 0 \\
\end{tabular}
\label{diquarkmodel}
\eeqn
With the choice of generalized GSO coefficients:
\beqn
&& c{b_j\choose \alpha,\beta,\gamma}=
   c{\alpha\choose 1}=
   c{\alpha\choose \beta}=\nonumber\\
&& c{\beta\choose 1}=
   c{\gamma\choose 1}=
  -c{\gamma\choose \alpha,\beta}=1\nonumber
\eeqn
$(j=1,2,3),$
with the others specified by modular invariance and space--time
supersymmetry. The sector ${\gamma}$ has the desired form
$\gamma_L^2=\gamma_R^2=4$ and give rise to exotic diquarks.

The classification by the type of exotics which
actually appear in specific models is crucial from a
phenomenological point view, as can be seen from the
following example. This is exemplified by examining the
exotic spectra in the models of refs. \cite{eu} and
\cite{custodial} and considering the neutrino
see--saw mechanism. Examining the exotic spectrum
in the model of ref. \cite{eu} we see that the
massless spectrum contains from the sectors
$b_2+b_3+\alpha\pm\gamma+(I)$ and
$b_1+b_3+\alpha\pm\gamma+(I)$ states which are
Standard Model singlets and carry $1/2$
the charge of the right--handed neutrino,
and it complex conjugate, with respect
to the $U(1)_{Z^\prime}$ which is embedded
in $SO(10)$. In The model of ref. \cite{eu}
we find that there exist such representations
which are $5$ and $\bar5$ of the hidden
$SU(5)$ gauge group, and some which are singlets
of the non--Abelian observable and hidden gauge
groups. The see--saw mechanism which operates
in this model makes use of a product of these
representations which forms the quantum numbers
of the complex conjugate of the right--handed
neutrino \cite{fhneut,fpnhmix}. Examining the
spectrum in the model of ref. \cite{custodial}
we observe that in this model all the exotic
states with similar $U(1)_{Z^\prime}$
charge, transform in representations of the
non--Abelian hidden gauge groups. Therefore
a similar combination of fields, which mimics
the charge of the conjugate of the right--handed
neutrino cannot be formed in this model.

\section{Classification by the hidden sector}

The hidden sector in the free fermionic models is
determined by the boundary condition of the
internal right-moving world--sheet fermions
${\bar\phi}^{1,\cdots,8}$. Several comments
are important to note. First we should define what is meant
by the hidden sector. The term hidden sector here means
that the states which are identified with the standard Model
states are singlets of the hidden sector. On the other hand,
both observable and hidden matter states may be charged with
respect to the flavor $U(1)$ symmetries and similarly the
exotic states may be charged with respect to the $U(1)$ symmetries
which are embedded in $SO(10)$ and may transform under the
non--Abelian gauge groups of the hidden sector. In the
models that I consider here there is no non--Abelian mixing,
{\it i.e.} there are no representations that transform
simultaneously under the non--Abelian gauge groups of the
observable and hidden sector. It should be however noted
that models with such representations may be constructed
as well.

At the level of the NAHE set the Hidden
gauge group is $E_8$. Let us recall that
the adjoint representation of $E_8$ decomposes as
$120\oplus128$ under $SO(16)$.
There are two sectors which produce the
adjoint representation of the $E_8$ gauge group.
The first is the Neveu--Schwarz sectors which produces
the 120 representation of $SO(16)$, and the second
is the sector ${\bf1}+b_1+b_2+b_3$, which produces
the spinorial 128 representation of $SO(16)$.
In addition to these two sources some models
may contain additional space time vectors
bosons which can enhance the hidden sector
gauge group. These additional space--time
vector bosons are obtained from combinations
of the boundary condition basis vectors which extend the
NAHE set.

The hidden sector gauge group depends on which of
these sectors contributes to the final hidden sector
gauge group. In the flipped $SU(5)$ and $SO(6)\times SO(4)$
models the hidden gauge group is typically broken
in two stages. The first is by the vector $2\gamma$
which breaks the $E_8$ gauge group to $SO(16)$
and the second stage is to an $SO(16)$ subgroup.
For example, in the revamped flipped $SU(5)$ model
the final hidden gauge group is $SO(10)\times SO(6)$.
In the flipped $SU(5)$ the vector $\gamma$ is the
basis vector which breaks $SO(10)\rightarrow SU(5)\times U(1)$
and has $1/2$ boundary conditions for
$\{{\bar\psi}^{1,\cdots,5},{\bar\eta}^1,{\bar\eta}^2,{\bar\eta}^3
{\bar\phi}^{1,\cdots,4}\}$. Recall from section (\ref{twozerosec})
that the vector $2\gamma$ is the basis vector that breaks
$(2,2)~\rightarrow~(2,0)$ world--sheet supersymmetry.
As we have seen in section (\ref{twozerosec}) this breaking
can also be achieved by a suitable choice of the GSO
projection coefficients. Therefore, the first stage
of the hidden sector gauge group breaking is the
breaking due to the breaking of the right--moving
$N=2$ world--sheet supersymmetry. This stage of
breaking is common to all the semi--realistic
free fermionic models which utilize the NAHE set.
The second stage of the breaking of the hidden sector
gauge group is more model dependent. The remaining
boundary conditions of the world--sheet fermions,
which determine the hidden sector gauge group, are to
a large degree fixed by the phenomenological requirements,
already imposed on the observable sector and by the
modular invariance constraints which constrain the
possible boundary conditions.

The space--time vector bosons from the Neveu--Schwarz
sector which contribute to the final hidden sector
gauge group are always present. The presence
of additional gauge multiplets from the sector
${\bf1}+b_1+b_2+b_3$, which enlarge the hidden sector
gauge group, depends on the GSO projection coefficients.
For example, in the model of ref. \cite{fny} the choice
$c{b_4\choose1}=1$ yields the projection of the
states from this sector and therefore the resulting
hidden sector gauge group arises solely from the
Neveu--Schwarz gauge multiplets. It should be
noted that the choice of GSO phases which
project the space--time vector bosons from the
sector ${\bf1}+b_1+b_2+b_3$ affect the
phenomenological properties of the observable sector.
The final hidden gauge group in a given model,
to some degree, is fixed due to the phenomenological properties
imposed on the observable sector.

An example which shows how this correlation works is
provided in the free fermionic standard--like models.
Suppose that we want to project all three
color triplets pairs from the Neveu--Schwarz sector.
It then follows that a vector $\alpha$, which breaks
the $SO(10)$ symmetry to $SO(6)\times SO(4)$ has an odd
number of periodic fermions from the set
$\{{\bar\psi}^{1,\cdots5},{\bar\eta}^1,{\bar\eta}^2,{\bar\eta}^3\}$.
To obey the modular invariance rule, $\alpha\cdot\gamma=0mod1$,
an odd number of fermions from the set $\{{\bar\phi}^{1,\cdots,8}\}$
must be periodic in the vector $\alpha$, and receive  boundary
condition of $1\over2$
in the vector $\gamma$. Therefore, the hidden gauge symmetry
is  broken in two stages. Typically it is broken to $SU(5)\times
SU(3)\times U(1)^2$. However, other possibilities do exist.
For example, modifying the vector $\gamma$ of ref. \cite{eu} by
$\gamma\{{\bar\phi}^3{\bar\phi}^4\}=1\rightarrow
\gamma\{{\bar\phi}^3{\bar\phi}^4\}=0$, enhances
the hidden gauge group from $SU(5)_H\times SU(3)_H\times U(1)^2$
to $SU(7)_H\times  U(1)^2$, for an appropriate choice of the generalized
GSO projection coefficients.

As noted above in some models additional space--time
vector bosons may arise from combinations of the
basis vectors which extend the NAHE set.
For some choices of the additional basis vectors that extend the NAHE
set, there exist a combination
\beq
X=n_\alpha\alpha+n_\beta\beta+n_\gamma\gamma
\label{enhancedgg}
\eeq
for which $X_L\cdot X_L=0$ and $X_R\cdot X_R\ne0$. Such a
combination may produce additional space--time vector
bosons, depending on the choice of GSO phases. For example, in the
model of ref. \cite{eu}
the combination $X=b_1+b_2+b_3+\alpha+\beta+\gamma$ has
$X_L\cdot X_L=0$ and $X_R\cdot X_R=8$. The space--time vector bosons
from this sector are projected out by the choice of GSO phases,
and this vector combination produces only space--time scalar
supermultiplets. However, with the modified GSO phases
in Eq. (\ref{euvectorbosons}) additional space--time
vector bosons are obtained from the sector
$b_1+b_2+b_3+\alpha+\beta+\gamma+(I)$ .
In this case the hidden $SU(3)_H$ gauge group is extended to
$SU(4)_H$.

Further classification of free fermionic models by the hidden sector
can be done by the hidden matter content. In all the
models the sector $b_j+2\gamma$ produce the vectorial 16
representation of the hidden $SO(16)$ gauge group decomposed
under the final hidden sector gauge group. These matter
representations are always present in the three generation models
that utilize the NAHE set of boundary condition basis vectors.
Additional hidden sector matter states arise from other combinations of
the additional basis vectors which extend
the NAHE set. These additional hidden matter play an important
role, for example, in generating a neutrino see saw mass matrix
\cite{fhneut}. In general, the number of additional hidden sector
matter states, transforming under the hidden sector non--Abelian
gauge groups, affects the scale at which the hidden sector
non--Abelian gauge groups become strongly interacting and therefore
is important, for example, in the context of supersymmetry breaking.

\section{Conclusion}

The realistic models in the free fermionic formulation have had
remarkable success in providing plausible explanations
to various properties of the Standard Model. Among those
we should list the natural emergence of three generations,
the qualitative structure of the fermion
mass spectrum and the possible resolution of the
string gauge coupling problem. Furthermore,
specific models can also explain the origin
of proton stability in a robust way.
The gross characteristics
of this class of models arise because of the
underlying $Z_2\times Z_2$ orbifold compactification at the
free fermionic point in the moduli space.
This compactification then leads to a large
number of three generation models, which differ by their
detailed phenomenological properties, giving rise
to the hope that a fully realistic model can be constructed.
While it is likely that superstring theory is only an
effective approximation to the truly fundamental Planck
scale theory, a fully realistic superstring model
is likely to be more than an accident.
Such phenomenological string models in turn will
serve as toy models in which we can learn about the
fundamental Planck scale dynamics.

In this paper I discussed in detail the basic ingredients and
building blocks that enter the construction of the realistic
free fermionic models. The aim  of this paper is to provide
some of the insight into the basic structures that underly these
models. The eventual goal of the program initiated here
is to uncover whether a fully realistic superstring
model in the free fermionic models can be constructed.
The classification of the models by boundary condition rules
was discussed. These rules illustrate how some
phenomenological properties of the models are fixed
by the boundary conditions.
The role of the free phases in
the determination of the physical properties of the
realistic string models was discussed in detail.
Mirror symmetry is seen to
arise in the fermionic models due to the discrete choices
of free phases. The mirror symmetry exist in these models
for the (2,2) models as well as for the more
general (2,0) models with periodic--antiperiodic
boundary conditions. Similar duality symmetries
can also be found in the extended models, which include
Wilson line breaking of the non--Abelian gauge symmetry.
Such duality symmetries provide one criteria for
classifying the models. The free phases also play an important
role in fixing the physical properties of the string models.
In this paper this role was illustrated in regard to the
final gauge group and the string gauge coupling unification
problem. The classification of the models by the pairing
of the world--sheet real fermions was elucidated,
and several examples demonstrated how the different
choices of pairings relate to the phenomenological
properties of the models. Further classification of
the models, by the different types of exotic representations
that may appear in the models and by the hidden sector
was discussed. Further classification of the models
by the anomalous $U(1)$ will be discussed in a separate
publication \cite{cf}.

%=============================================================================
%=============================================================================
\bigskip
\medskip
\leftline{\large\bf Acknowledgments}
\medskip

I would like to thank the Weizmann Institute and CERN theory group
for hospitality while part of this work was conducted.
This work was supported in part by the Department of Energy
under Grant No.\ DE-FG-0586ER40272.

%=========================================================================
%======================== REFERENCES =====================================
%=========================================================================

\vfill\eject

\bigskip
\medskip

\bibliographystyle{unsrt}

\vfill\eject
\end{document}